\documentclass[11pt]{article}
\usepackage{graphicx}
\usepackage{natbib}
\usepackage{amsmath,amsfonts,amssymb,amsbsy,amsthm,mathtools}
\usepackage{bbm,bm, centernot}
\usepackage{color}
\usepackage{subfig,float}
\usepackage{booktabs}
\usepackage{url}
\usepackage[a4paper]{geometry}
\allowdisplaybreaks
\numberwithin{equation}{section}
\usepackage{lmodern}
\usepackage{paralist}

\newcommand{\Prob}{\textnormal{P}}

\newcommand{\suit}[1]{\left(#1\right)}

\title{Tail Risk Analysis for Financial Time Series}

\author{Anna Kiriliouk\thanks{Namur Institute for Complex Systems, University of Namur, Rue Graf\'{e} 2, 5000 Namur, Belgium. E-mail: anna.kiriliouk@unamur.be}
\and Chen Zhou\thanks{Erasmus School of Economics, Erasmus University of Rotterdam, Burgemeester Oudlaan 50, 3062 PA Rotterdam, the Netherlands. E-mail: zhou@ese.eur.nl}}

\begin{document}
\maketitle

\begin{abstract}
This book chapter illustrates how to apply extreme value statistics to financial time series data. Such data often exhibits strong serial dependence, which complicates assessment of tail risks. We discuss the two main approches to tail risk estimation, unconditional and conditional quantile forecasting. We use the S\&P 500 index as a case study to assess serial (extremal) dependence, perform an unconditional and conditional risk analysis, and apply backtesting methods. Additionally, the chapter explores the impact of serial dependence on multivariate tail dependence.
\end{abstract}

\section{Introduction}
Tail risk analysis has increasingly become a focal point in financial risk management. For instance, in the Basel Framework, the primary global standard setter for the prudential regulation of banks, capital requirements for market risk are directly associated with tail risk measures for banks' profits and losses such as the Value-at-Risk (VaR) and the Expected Shortfall (ES). As per Basel III, the VaR is a measure of the worst expected loss at a pre-defined confidence level. Since the pre-defined confidence level is often set to close to 1, statistically speaking, the VaR is a high quantile of the losses, i.e., an extreme risk measure. 

To set a reliable capital requirement, robust and efficient methods for assessing extreme risks are required by both the financial industry and regulatory bodies. In this context, extreme value statistics emerge as a powerful tool. They offer a means for extrapolation from available data to extreme events, which is particularly relevant for risk analysis in scenarios with limited data available on extreme events.

However, applying extreme value statistics to financial data presents a significant challenge due to the strong serial dependence often observed in such data. Such serial dependence can lead to potentially inaccurate risk assessments. This chapter aims to showcase various methodologies for applying extreme value statistics to financial data. We place a special emphasis on addressing and overcoming the challenges posed by serial dependence, aiming to provide more accurate and reliable tools for tail risk analysis in the financial sector. Note that Chapter~14 assesses the influence of serial dependence on the analysis of climate data. Our case studies show that the serial dependence in financial data is more persistent after applying typical declustering methods used for climate data, thus more challenging to handle.

Studies applying extreme value statistics to financial data fall into two categories, each with its own approach and interpretation.

The first stream of literature focuses on the use of raw financial data, recognizing serial dependence in the data. In particular, \cite{drees2003extreme} demonstrates that classical extreme value estimators based on independent and identically distributed (i.i.d.) observations, such as the Weissman estimator for high quantiles \citep{weissman1978estimation}, are still valid under weak serial dependence. The main difference is that the asymptotic variance of such estimators can be inflated due to the presence of serial dependence. In \cite{de2016adapting}, it is shown that the asymptotic bias in such estimators can be effectively corrected at the cost of a further inflated asymptotic variance, again related to serial dependence. Even though the asymptotic variance associated with this estimator can be consistently estimated, it often results in wider confidence intervals for unconditional quantiles.

Using the raw data results in an estimation of unconditional quantiles of the financial time series. This is primarily associated with regulatory perspectives, where the goal is to establish capital requirements, absorbing potential losses in extreme events. For regulatory purposes, a stable and robust risk measure is needed to establish a long-run capital requirement. The risk measure should account for all potential macroeconomic scenarios. Unconditional quantiles provide such a broad view of risk, regardless of the underlying economic situation; therefore, they are crucial for setting baseline regulatory standards. This regulatory perspective is fundamental for ensuring the overall stability of financial institutions by mandating capital buffers that can withstand extreme market movements.

By contrast, the second stream of literature involves using a financial time series model first, followed by the application of extreme value statistics to the residuals. \cite{mcneil2000estimation} follow this approach to quantify the conditional quantile of future losses given all information available at a certain time in history, while \cite{hoga2019confidence} provides the theoretical justification, showing that the unconditional quantile estimator based on such a two-step approach still possesses asymptotic normality.

The two-step approach, focusing on conditional quantiles, is aligned with the perspective of risk monitoring. Due to the incorporation of most recent information available, often the conditional volatility of the underlying time series, the conditional quantile adapts to the conditional distribution of the losses and therefore can fluctuate alongside other distributional parameters such as the conditional volatility. In other words, it accounts for the underlying macroeconomic scenario and provides a more timely risk forecast. Estimating conditional quantiles is instrumental in understanding the potential extent of losses or how adversely a financial position can deviate under the ongoing market conditions. This method is particularly relevant for internal risk management within financial institutions, where dynamic assessments of risk are crucial for day-to-day operations and strategic decision-making.

To summarize, the two different approaches have different application perspectives when applied to financial data. The unconditional quantile focuses on the overall adequacy of the capital buffer, while the conditional quantile approach delves into the dynamic nature of risk and its day-to-day management. They are integral to comprehensive risk management. For instance, in setting the countercyclical buffer, regulators also need to consider conditional quantiles, acknowledging that risk levels can vary significantly over time and across different market conditions. This reconciliation underlines the importance of both approaches for a comprehensive financial risk assessment and management.

Choosing between these two approaches is also related to choosing backtesting methods. Backtesting methods use new observations to verify the risk forecasts. For instance, the coverage tests focus on whether the predicted quantile covers the realized losses at the intended probability level. When using unconditional quantiles, the focus is primarily on coverage tests for a long horizon, assessing the accuracy of the risk models in capturing extreme events under different market scenarios. However, with conditional quantiles, if they are well predicted, events where realized losses exceed the predicted quantiles should be independent over time. This difference underlines the varied complexities and considerations inherent to each approach.

Estimating financial risk measures such as the VaR is a univariate statistical problem, although the univariate series might possess serial dependence. In an orthogonal direction, one may also consider multivariate analysis, i.e., analyzing multivariate tail events defined by several risk factors. In the context of financial risk analysis, this is typically referred to as measuring systemic risk. Multivariate extreme value statistics provide tools for that purpose. When applying existing methods in multivariate extreme value statistics to financial data, serial dependence in marginal times series may pose a distortion in estimation as well. For that purpose, this chapter also examines the impact of serial dependence on estimating multivariate tail dependence measures.  

In the next sections, we employ the S\&P 500 index as an example and perform both unconditional and conditional quantile forecasts using available techniques in extreme value statistics. Section \ref{sec:explore} shows some exploratory analysis to uncover the properties of financial data for further modeling, namely heavy-tailedness and serial dependence. We also discuss potential declustering methods and examine their effectiveness in reducing serial dependence. In Section \ref{sec:unconditional}, we conduct risk forecasts for unconditional quantiles, while showing the conditional risk analysis in Section \ref{sec:conditional}. Section \ref{sec:backtesting} introduces the backtesting methods and backtests the unconditional and conditional risk forecasts. In Section \ref{sec:multivariate}, we examine the impact of serial dependence on estimating multivariate tail dependence measures. Section \ref{sec:conclusion} concludes the chapter. Finally, some (historical) references and comments are consolidated in Section \ref{sec:notescomments}.

\section{Exploratory Analysis} \label{sec:explore}
%\cz{Done: Add a brief sentence to clarify the fundamental role of the S\&P 500 index in the world’s financial markets}.
We take the S\&P 500 index as an example dataset. The S\&P 500 tracks the stock performance of 500 of the largest companies listed on stock exchanges in the United States. They include approximately 80\% of the total market capitalization of U.S. public companies.

We take the daily adjusted closing prices of the S\&P 500 index from Yahoo Finance, for the period January 1st, 1961 up to December 31st, 2022, denoted $P_0,\ldots,P_{n}$. The negative daily log-returns, denoted by $X_1,\ldots,X_n$, are defined via
\[
X_i = - \log \left( P_i / P_{i-1} \right), \qquad i =1,\ldots,n.
\]
Here we use the log-returns since they are regarded as continuously compounded returns of this series and can take values in the entire real line. Nevertheless, negative daily log-returns can be regarded as a proxy for daily losses: given that $- \log y$ behaves like $1 - y$ for $y$ near $1$, whenever $P_i \approx P_{i-1}$, we can set $y = P_i/P_{i-1}$ and obtain that $X_i \approx \left( P_{i-1} - P_i \right)/P_{i-1}$. 

\begin{figure}[ht]
\centering
\includegraphics[width=0.9\textwidth]{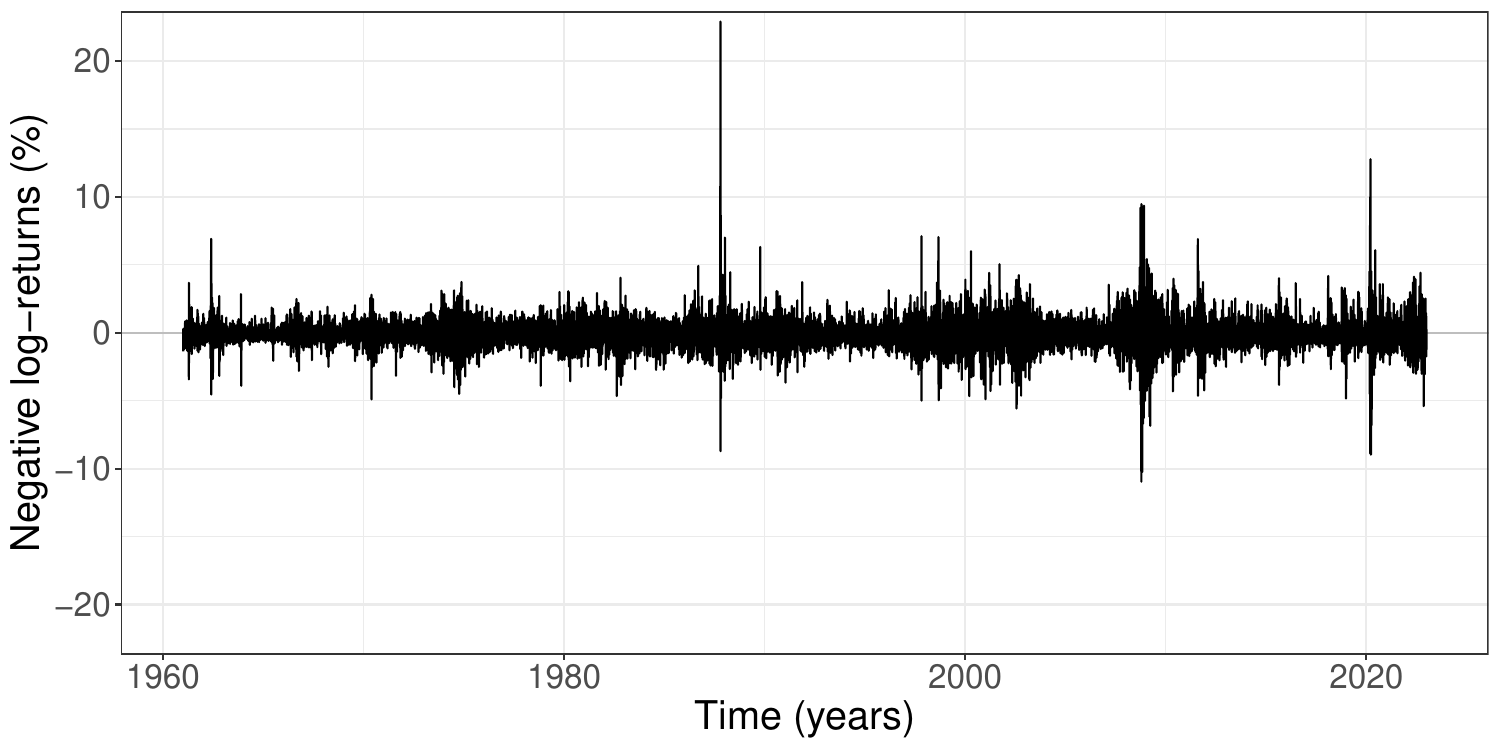} 
\caption{Daily negative log-returns of the S\&P 500, for the period January 1st, 1961 up to December 31st, 2022.}
\label{fig:sp500}
\end{figure}

For the S\&P 500 returns in Figure~\ref{fig:sp500} (sample size $n = 15 \, 605$), we observe a clear volatility clustering effect in the series as well as clustering for extremes: there is a tendency for extreme returns to be followed by other extreme returns. In the following, the negative daily log-returns will be denoted by $X_1,\ldots,X_n$. Assuming that the underlying distribution is stationary, we use $F$ to denote the stationary unconditional distribution.

\subsection{Heavy-tailedness}
\begin{figure}[ht]
\centering
\includegraphics[width=0.45\textwidth]{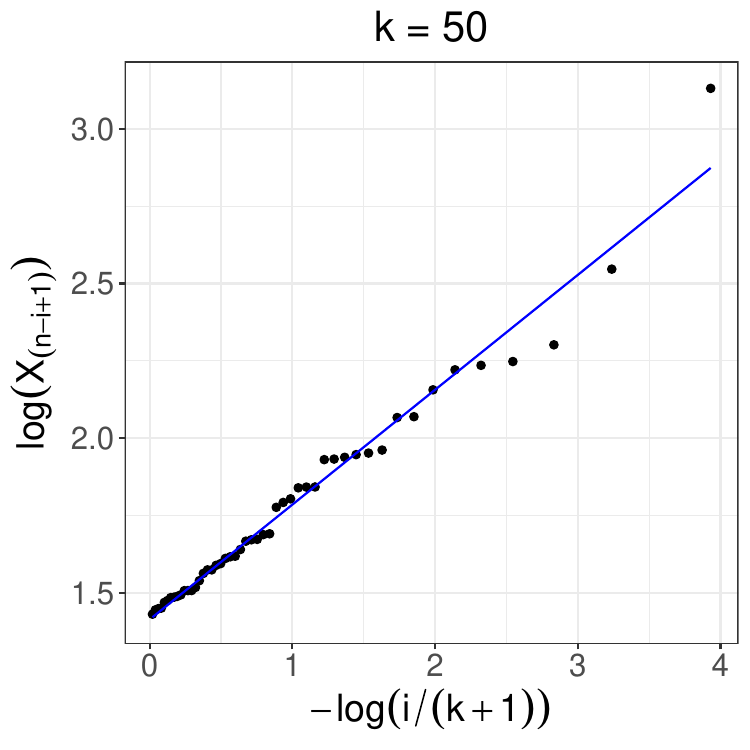} 
\includegraphics[width=0.45\textwidth]{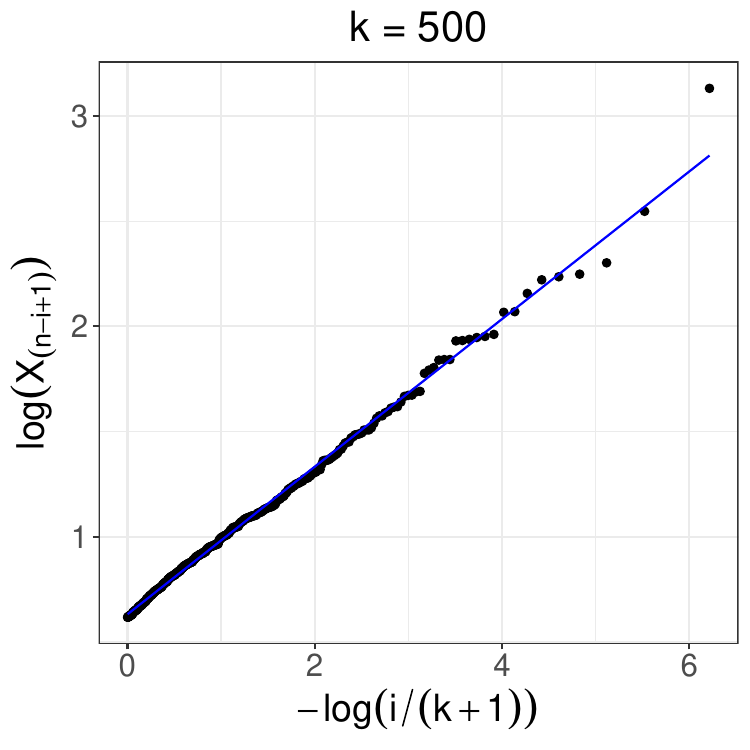} 
\caption{Pareto quantile plots for the daily negative log-returns of the S\&P 500 for $k = 50$ and $k = 500$.}
\label{fig:sp500par}
\end{figure}

To explore the tail of the distribution of the loss returns, we investigate the Pareto quantile plots for $k = 50$ and $k = 500$ in 
Figure~\ref{fig:sp500par}, i.e., we plot
\[
\left\{ \left( - \log \left( \frac{i}{k+1} \right), \, \log X_{(n-i+1)} \right) \right\}_{i=1}^k, 
\]
where $X_{(1)}\leq  \cdots \leq X_{(n)}$ are the order statistics of the observations. The figure shows a clear linear relation between $-\log \{1-F(x)\}$ and $\log x$ for sufficiently large $x$. By writing this linear relation as
$$-\log \{ 1-F(x)\} \approx \alpha \log x + b,$$
we obtain an approximate power law in the tail: for large $x$, $1-F(x)\approx Ax^{-\alpha}$, where $A=e^{-b}$. The power law resembles the definition of heavy tails as stated in \eqref{eq:regvar}. Referring to this formal definition, the parameter $\alpha$ is the \emph{tail index}. 

From Figure~\ref{fig:sp500par}, we obtain an initial estimator for $\alpha$ by inverting the slope of the regression line; we find $\widehat{\alpha} = 2.69$ for $k = 50$ and $\widehat{\alpha} = 2.85$ for $k = 500$. Note that the linear relation in the log-scaled scatter plot between tail probabilities and quantiles is applicable to not only the S\&P 500 index, but to most financial loss returns such as stocks, bonds and exchange rates. After estimation of the slope, this linear relation is the foundation of extrapolating from observed quantiles to extreme quantiles.

%\cz{Done: explaining that this is a general phenomenon and its implication for risk analysis.}

\subsection{Serial Dependence}
We first perform an Augmented Dickey--Fuller (ADF) test to test the null hypothesis of having a unit root, i.e., whether the lagged values $X_{i-1}$ are irrelevant in predicting the change in $X_{i}$, $i=2,\ldots,n$. With a p-value of virtually zero, we reject the null of having a unit root and thus regard the time series of the S\&P 500 returns as stationary.

\begin{figure}[p]
\includegraphics[width=0.9\textwidth]{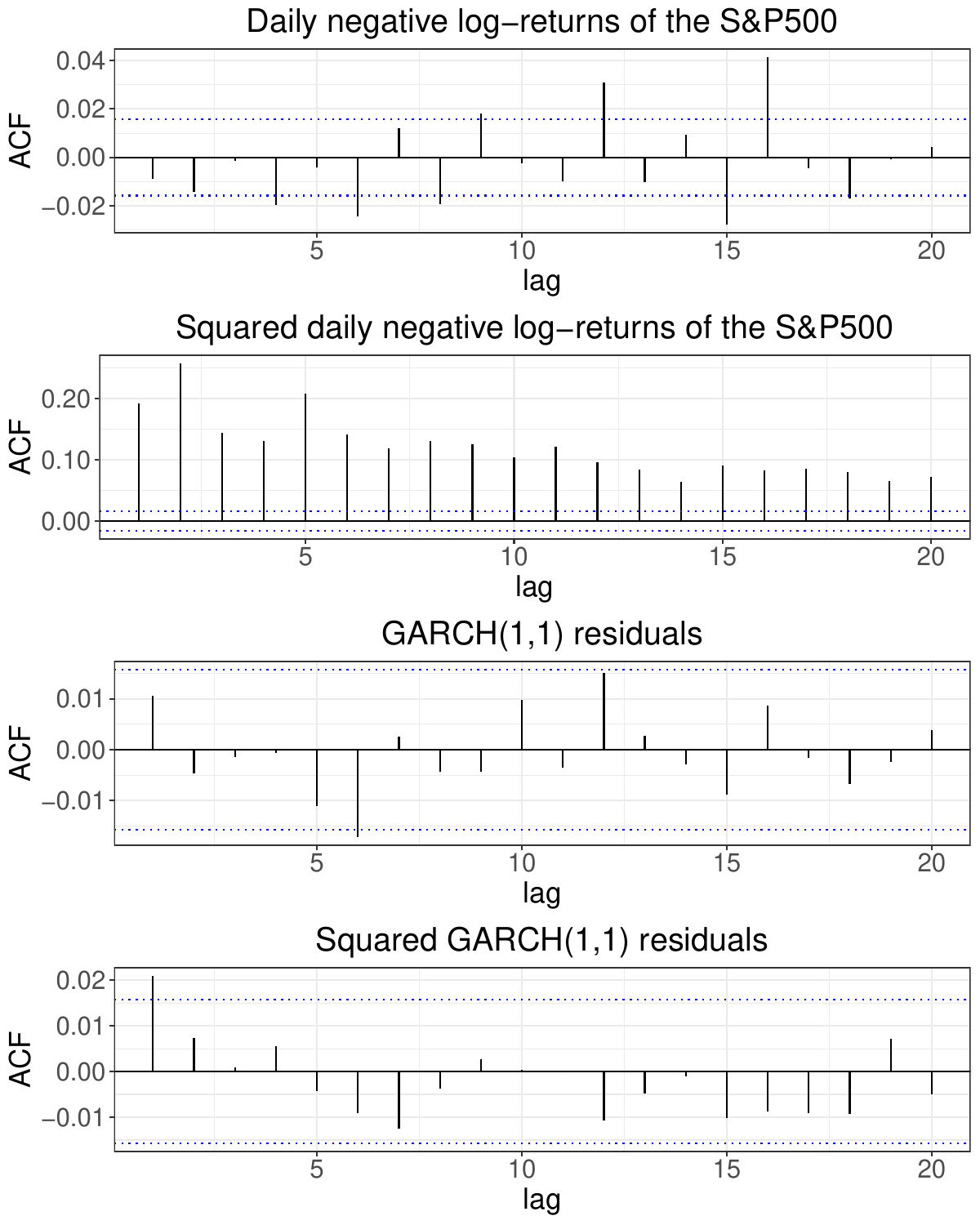}  
\caption{Autocorrelation function for the original series of daily negative log-returns and their squared values (top) and for the residuals $\varepsilon_t$ of the AR(1)--GARCH(1,1) model and their squared values (bottom).}
\label{fig:sp500acf}
\end{figure}

To investigate the serial dependence in this dataset, Figure~\ref{fig:sp500acf} shows the autocorrelation up to lag 20 for the daily negative log-returns and the squared daily negative log-returns (top two panels). Clearly, there is strong serial dependence for the squared returns, and some serial dependence for the returns itself.

Next, we fit an  AR(1)--GARCH(1,1) model \citep{bollerslev1986generalized} to these observations as follows:
        \[
X_t = \mu + \phi X_{t-1} + \sigma_t \varepsilon_t, \qquad \sigma_t^2 = \omega + \alpha \varepsilon_{t-1}^2 + \beta \sigma^2_{t-1},
    \]
where the residuals $\varepsilon_t$ are assumed to be well-behaved i.i.d. random variables for $t=2,\ldots,n$. Here we do not make any distributional assumption for the residuals in our further analysis. The parameters $\mu$, $\phi$, $\omega$, $\alpha$ and $\beta$ can be estimated using quasi-maximum likelihood estimation (QMLE), i.e., a conditional likelihood function is constructed by assuming that  $\varepsilon_t$ follow an i.i.d. standard normal distribution. Note that the normality assumption is only used for estimating the parameters. The QMLE results in estimates that still possess consistency and asymptotic normality under mild conditions; see, e.g. \cite{hall2003inference}. After estimating the parameters, we filter out the volatility series $\{\hat \sigma_t\}_{t=1}^n$ and the residual series $\{\hat \varepsilon_t\}_{t=1}^n$ and further model the residuals. 

By estimating the model using QMLE, we find the following point estimates (with standard deviations in parentheses): $\hat{\mu} = -0.050$ $(0.0064)$, $\hat{\phi} = 0.066$ $(0.0098)$, $\hat{\omega} = 0.011$ $(0.003)$, $\hat{\alpha} = 0.099$ $(0.0135)$, and $\hat{\beta} = 0.894$ $(0.0133)$. Note that $\hat{\alpha}+\hat{\beta}<1$ but is close to 1, therefore, the time series is close to violating the assumption of covariance stationarity.

We further investigate the estimated residuals after filtering by the AR(1)--GARCH(1,1) model. Performing an ADF test again rejects the presence of a unit root with a p-value that is virtually zero. In addition, Figure \ref{fig:sp500acf} shows the corresponding autocorrelation plots (bottom two panels). Different from the original series, neither the original nor the squared residuals show any serial correlation. These figures provide empirical evidence that after filtering, the residuals possess weak serial dependence and can be regarded as serially independent.

\subsection{Serial Dependence in Extremes}
While we use the autocorrelation to measure serial dependence in observations at moderate level, it is also important to measure serial dependence in extreme events. For that purpose, we employ the \emph{extremal index} \cite{leadbetter1983extremes}, the most well-known measure capturing serial dependence across the extremes of a time series; see, also Chapter~14.

The extremal index for a stationary time series $X_1,X_2,X_3,\ldots$ with distribution function $F$ is defined as follows. Let $M_{a,b} = \max (X_{a},...,X_b)$ for $a, b \in \mathbb{N}$, $a < b$. If, for each $\tau$, there exists a sequence $\{u_n(\tau)\}$ such that $$\lim_{n \to \infty}  n \left\{1-F(u_n)\right\} = \tau, \qquad  \lim_{n \to \infty} P(M_{1,n} \leq u_n) = e^{-\theta \tau},$$
then $\theta$ is called the \emph{extremal index} of the times series. Note that for an i.i.d. sequence, we have $\theta=1$. By contrast, for a duplicated sequence $X_1,X_1,X_2,X_2,X_3,X_3,\ldots$ where $X_1,X_2,X_3,\ldots$ are i.i.d., the extremal index is $\theta=1/2$. More generally, for a time series where i.i.d. observations are duplicated $m$ times, we have $\theta=1/m$. Conversely, the quantity $1/\theta$ reflects the potential cluster size of extreme events in a time series. Intuitively, if extreme observations are more than $1/\theta$ apart, they can be considered as independent events. This motivates a declustering procedure creating gaps of length at least $1/\theta$, for which estimating $\theta$ is a crucial first step.

We estimate the extremal index using the sliding block estimator \citep{northrop2015,berghaus2018weak}. Despite being computationally simple and depending on only one tuning parameter (the block size), it has been shown to perform very well compared to its competitors. The idea is as follows. First, consider the marginal transformation to standard uniform, $U_i=F(X_i)$ for $i \in \mathbb{N}$. %Then $U_i$ follows the uniform distribution on $[0,1]$. 
Denote $N_{1,b}=F(M_{1,b})=\max(U_1,\ldots, U_b)$. Then, as $b\to\infty$,
\[
\Prob(-b\log N_{1,b}\geq x)\to e^{-\theta x}.
\]
In other words, for a sufficiently large block size $b$, $-b\log N_{1,b}$ follows an approximate exponential distribution with mean $1/\theta$.

This intuition leads to the following estimator. Denote $n=b_nk_n$, where $n$ is the sample size and $b=b_n$ is the block size. Let $\hat F_n(x)= n^{-1} \sum_{i=1}^n \mathbb{I} \suit{X_i\leq x}$ be the empirical distribution function and define the pseudo-observations $\hat Y_i= - b_n\log \hat F_n\suit{M_{i,i+b_n}}$, for $i=1,\ldots, n-b_n$.  Based on $\hat Y_{i}$, we can estimate the extremal index by
 $$\hat \theta = \suit{ \frac{1}{n-b_n} \sum_{i=1}^{n-b_n}{\hat Y_i}}^{-1}.$$
 We refer to this estimator as the \emph{bias-corrected sliding blocks estimator}.

\begin{figure}[ht]
    \centering
    \includegraphics[width=0.7\textwidth]{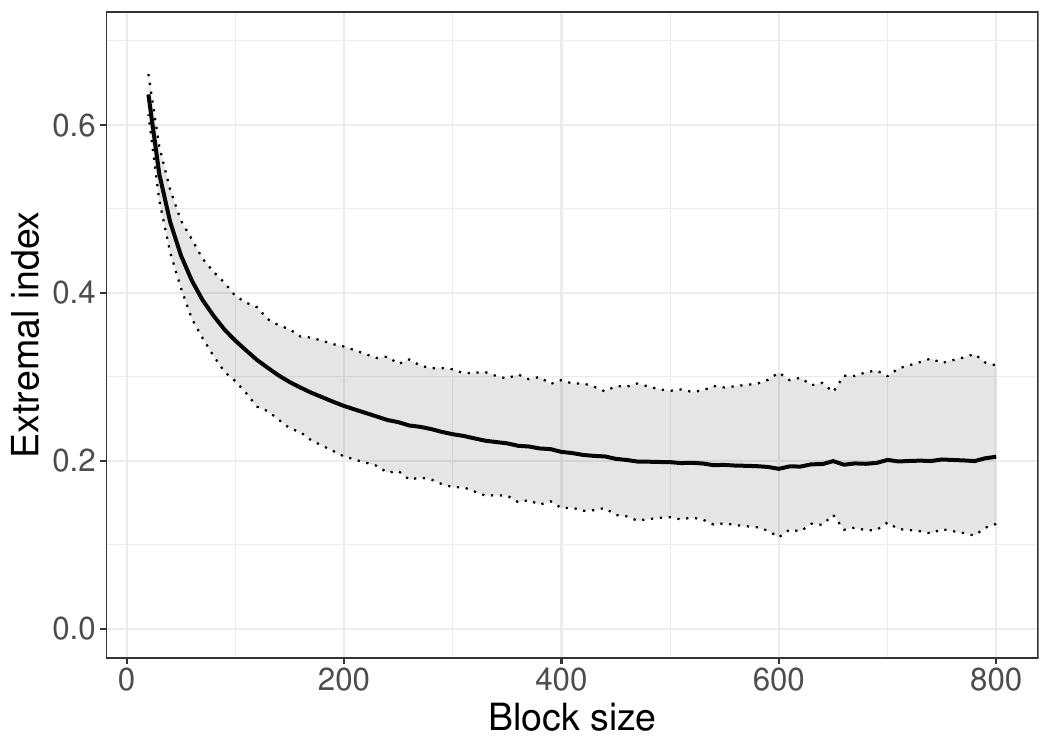} 
    \caption{Bias-corrected sliding blocks estimator, with 95\% confidence intervals, for the daily negative log-returns of the S\&P 500.}
    \label{fig:sp500theta}
    \end{figure}

We use the \texttt{exdex R} package \citep{exdex} to estimate the extremal index for the S\&P 500 daily negative log-returns. Figure~\ref{fig:sp500theta} shows the bias-corrected sliding block estimator, with 95\% confidence intervals based on an adjustment of a naive (pseudo-)loglikelihood \citep{northrop2015} for various block sizes. We choose a block size of $b = 500$ in the estimation, since the point estimates stabilize from that point on. This results in a final estimate of $0.198$ with 95\% confidence interval $(0.133, 0.283)$. In other words, the (limiting) average cluster size for the extremes is roughly between $4$ and $6$.

We use the estimated extremal index to guide the following declustering procedure, aiming to remove the serial dependence in the extremes. We compare two techniques for declustering:
\begin{enumerate}
    \item We take every fifth observation, i.e., all Monday negative log-returns. For completeness, we can do the same for the other weekdays, Tuesday to Friday. This leads to five series of declustered data of sizes $n = 2977$, $3189$, $3179$, $3140$ and $3120$ respectively.
    \item We first order all \emph{positive} observations (negative log-returns) from highest to lowest, and start by selecting the two largest observations. 
    If the second largest observation was recorded within two consecutive days of the first, then it is removed; otherwise both days are kept. For subsequent ordered values, we remove the corresponding day if it is recorded within two consecutive days of any of the previously kept days. We apply the same procedure to the \emph{negative} observations. 
\end{enumerate}

The first procedure is simple but it does \emph{not} explicitly focus on extremes. In other words, some extreme observations will be missed due to this arbitrary procedure. By contrast, the second procedure retains extreme observations as much as possible, while still creating a gap between them. Following the theoretical discussion on the extremal index, we expect that the selected subsample (or declustered subsample) will possess weak serial dependence, at least in the extremes.

We remark that both declustering methods are used in the literature, for instance, when dealing with serial dependence in climate data; see, e.g., \cite{einmahl2022spatial}, which uses the second procedure. For climate data, such declustering methods are successful in eliminating serial dependence in the data, at least for extreme observations: the declustered subsamples often possess an extremal index close to 1 and statistical tests cannot reject the null hypothesis that the extremal index is 1. Even though $\theta=1$ does not necessarily correspond to an i.i.d. sequence, the weak evidence that $\theta=1$ is not rejected is often used as justification for these declustering procedures.

To verify whether such procedures can eliminate serial dependence in extremes for financial data, we re-estimate the extremal index of the declustered subsample.

For the first procedure, we estimate the extremal index for each subsample based on one particular weekday. We observe that the subsamples exhibit slightly lower serial dependence in the extremes, but the estimates of $\theta$ are still far from $1$. For example, for a block size $b = 200$, which gives an approximately stable region for all five datasets, we find the following point estimates and $95\%$ corresponding confidence intervals for the five subsamples: $0.495$ $(0.334,0.703)$, $0.389$ $(0.229,0.613)$, $0.359$ $(0.244,0.505)$, $0.291$ $(0.216,0.382)$ and $0.353$ $(0.247,0.485)$ respectively. The conclusion is that the declustered subsamples possess extremal indices significantly below 1.

The second procedure is an adaptation of the one in \cite{einmahl2022spatial}. A dataset obtained in such a manner is expected to show less serial dependence because all retained extreme observations have also a gap of $1/\theta$ periods. The declustering is applied to both positive and negative observations to avoid over-representing one of the two tails, leading to a dataset appropriate for quantile estimation (as in the next section). %Note that if the declustering procedure is to be used for, say, quantile esitmation as in the next section, we should apply the same procedure to the \emph{negative} observations, so that the left tail will not be over-represented.

After declustering, we again re-estimate the extremal index for the retained observations. Surprisingly, estimates for $\theta$ remain small; taking $b = 250$ (for a sample size of $n = 7053$) gives a point estimate of $0.247$ with a 95\% confidence interval $(0.165,0.352)$. Note that in the described procedure, we create a gap of three by excluding extreme events within two consecutive days. Increasing the gap from two consecutive days to nine consecutive days (i.e. a gap of 10 days, equivalent to two trading weeks), we find a sample of size $n = 2360$ and $\hat{\theta} = 0.369$ with 95\% confidence interval $(0.249, 0.522)$ for $b= 150$. This result is similar to the estimated extremal indices of the first procedure. Increasing the range further to twenty consecutive days (i.e. a trading month), we find a sample of size $n = 1155$ and $\hat{\theta} = 0.437$ with 95\% confidence interval  $(0.319, 0.582)$ for $b= 100$. Again, this declustering procedure leads to a declustered subsample with an extremal index significantly below 1.

Although $\theta=1$ does not necessarily correspond to an i.i.d. sequence, having $\theta<1$ rejects the null of serial independence, even for extreme values. To conclude, all proposed declustering procedures fail to eliminate serial dependence in the extreme values of this financial dataset.

We investigate this phenomenon further via a small simulation study. We simulate from an AR(1)-GARCH(1,1) model with the same parameters as the point estimates obtained above when using the full sample ($n = 15 \, 605$). For each simulated dataset of size $n$, we first estimate its extremal index. By repeating this procedure for 100 simulated samples, we find an average value of $\hat{\theta} = 0.212$ for $b = 500$, close to what we have obtained from the real data.
Next, we apply the first declustering procedure (taking only every fifth observation) to each sample, and then estimate the extremal indices for each declustered subsample. Across the 100 simulated subsamples, we obtain an average of $\widehat{\theta} = 0.459$. Similarly, by applying the second declustering procedure and taking the average of the estimated extremal indices for each declustered subsample, we obtain average of $\widehat{\theta} = 0.262$ (for range 2) and $\widehat{\theta} = 0.372$ (for range 9). These values are in line with what we found for the S\&P 500 data. We conclude that the AR(1)-GARCH(1,1) type of serial dependence creates a difficult challenge for the declustering procedures motivated by the extremal index. This is a genuine problem in financial time series model, which worthies further investigation.

\section{Unconditional Risk Analysis}\label{sec:unconditional}
In this section, we conduct  unconditional risk analysis aiming to estimate the 99\% quantile of the stationary distribution, based on the original dataset. We apply the high quantile estimator in \cite{weissman1978estimation}, albeit with caution that the data are serially dependent. 

Consider the observed time series $X_1,\ldots,X_n$ possessing potential serial dependence. %Denote the stationary distribution of the time series as $F$. 
Assume that the underlying stationary distribution $F$ has a regularly varying tail, i.e.,
\begin{equation}\label{eq:regvar}
\lim_{t\to\infty}\frac{1-F(tx)}{1-F(t)}=x^{-\alpha},
\end{equation}
where $\alpha > 0$ is the tail index. Examples of such heavy-tailed distributions are the Pareto distribution, the Student-$t$ distribution, and the Fr\'{e}chet distribution. The goal is to estimate $F^{-1} (p)$, the high quantile of the distribution function $F$, for some $p=p_n$ with $p_n \rightarrow 1$ as $n\to\infty$.

Let $k = k_n$ denote an intermediate sequence such that, as $n \rightarrow \infty$, $k \rightarrow \infty$ and $k/n \rightarrow 0$. Let $X_{(1)} \leq \cdots \leq X_{(n)}$ be the order statistics of $X_1,\ldots,X_n$. Let $\widehat{\alpha}$ denote an appropriate estimator of the tail index $\alpha$. Then, the Weissman estimator \citep{weissman1978estimation} is defined as
\begin{equation}\label{eq:quant1}
    \widehat{F}^{-1}(p) = X_{(n-k)} \left\{ \frac{k}{n(1-p)} \right\}^{1/\widehat{\alpha}}.
\end{equation}

To apply the Weissman estimator in \eqref{eq:quant1} to the S\&P 500 data, we proceed as follows. Firstly, we need to estimate $\alpha$; we compare the standard Hill estimator (see, for example, Chapter~6) with the bias-corrected Hill estimator for serially dependent data proposed in \cite{de2016adapting}. 
In the bias-corrected Hill estimator, one needs to estimate a second order parameter (for details, see \cite{de2016adapting}). Here we adopt a pragmatic approach by fixing this parameter to a constant value $-1$.\footnote{The choice of this specific value is motivated by a boundary condition when comparing the performance of the peaks-over-threshold method and the block maxima method. For more details on the comparison, see \cite{bucher2021horse}. We also estimated the second order parameter $\rho$ as proposed in \cite{de2016adapting}, leading to $\hat{\rho} = -1.02$ for $k_{\rho} = 1250$. The estimate is very close to the choice in the pragmatic approach.} 

%In the empirical analysis below, we shall use a block bootstrapping method to approximate the confidence interval of the estimators.

\begin{figure}[ht]
    \centering
    \includegraphics[width=0.45\textwidth]{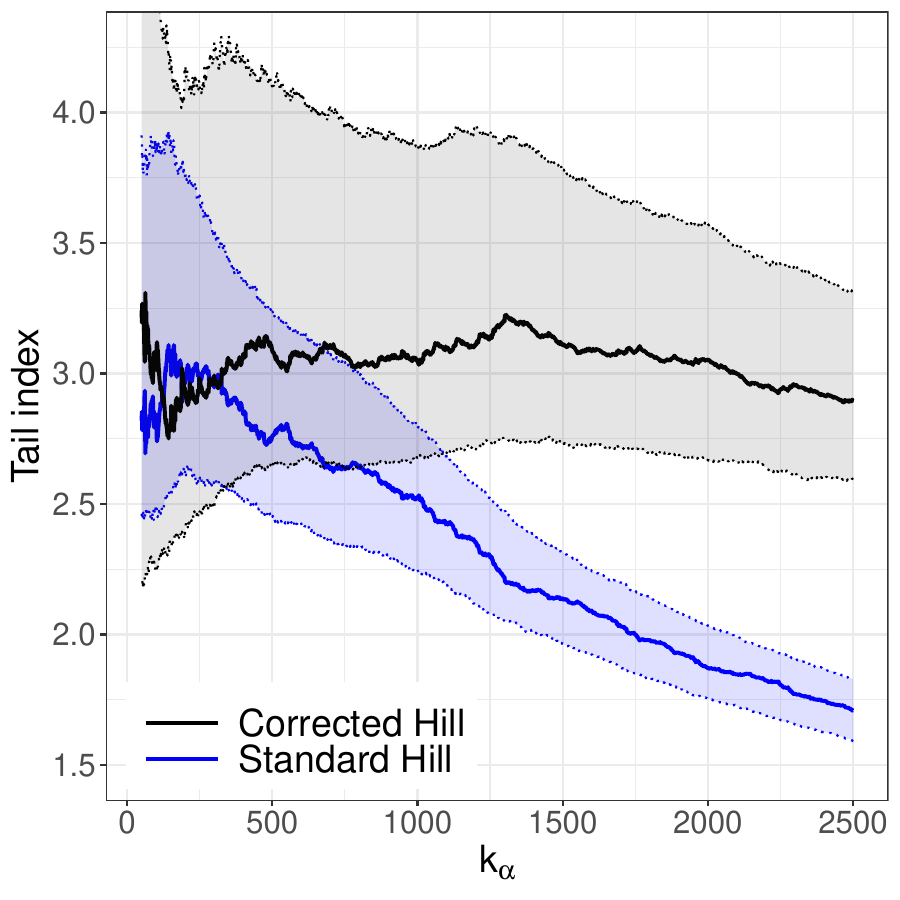} 
    \includegraphics[width=0.45\textwidth]{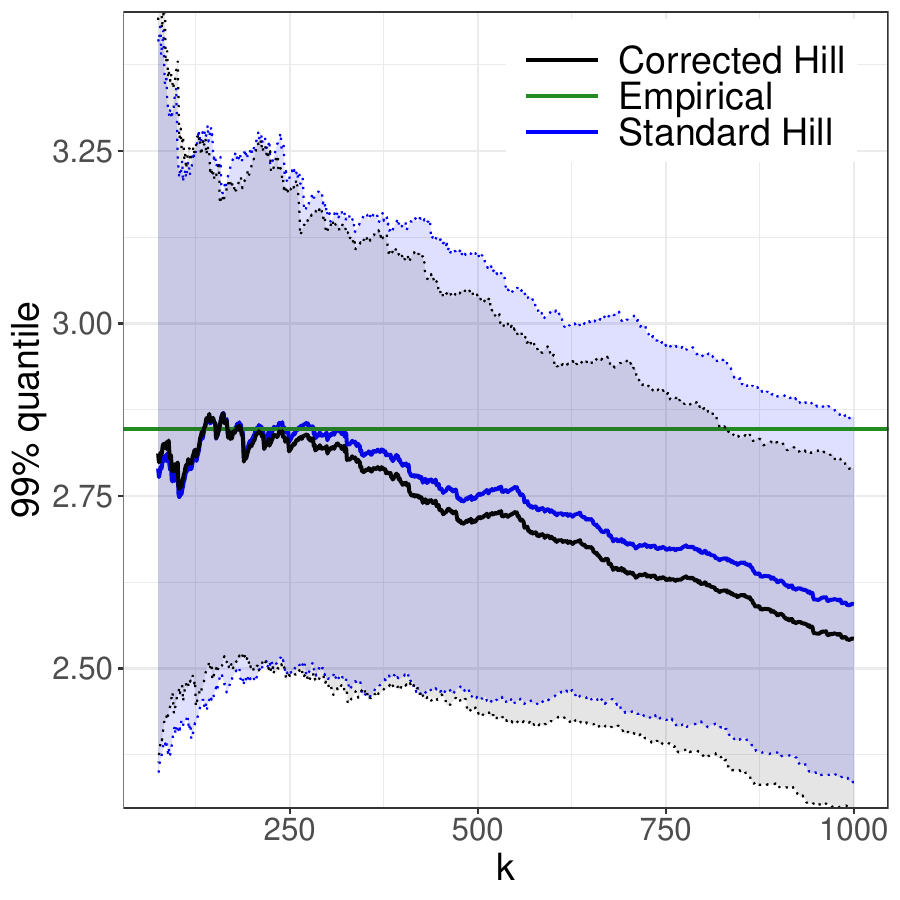} 
    \caption{Estimates of the tail index $\alpha$ and the 99\% quantile based on the standard Hill estimator (blue line) and bias-corrected Hill estimator (black line) for the daily negative log-returns of the S\&P 500. The horizontal green line represents the empirical 99\% quantile.}
    \label{fig:sp500hill}
\end{figure}

Figure~\ref{fig:sp500hill} (left) shows the results for both approaches, with the 90\% confidence intervals obtained via a block bootstrap. We use a block bootstrap with a random block length following a geometric distribution with mean $200$. Note that estimating $\alpha$ requires choosing a suitable value for $k_{\alpha}$. We observe that the bias-corrected Hill estimator allows for a wider and higher choice of $k_{\alpha}$, without imposing a bias to the estimate. Based on the standard Hill estimator, choosing $k_{\alpha} = 250$ leads to an estimate $\widehat{\alpha} = 2.95$. By choosing $k_{\alpha} = 1000$ for the bias-corrected Hill estimator, we obtain a similar estimate $\widehat{\alpha}^{(c)} = 3.05$.

Next, we plot the quantile estimator \eqref{eq:quant1} as a function of $k$ in Figure~\ref{fig:sp500hill} (right). Here we plug in the estimates of $\alpha$ obtained in both  approaches without varying $k$ further. %via the corrected Hill estimator (with $\rho = -1$)  $\widehat{\alpha}^{(c,-1)} = 3.051$ for $k = 750$, the corrected Hill estimator with estimated $\rho$, $\widehat{\alpha}^{(c,-1)} = 3.051$ for $k = 750$, and the standard hill estimator,   $\widehat{\alpha} = 3.051$ for $k = 250$. 
The horizontal line indicates the empirical 99\% quantile estimate at $2.85$. Given the large sample size of the full dataset, an empirical estimate is reliable and serves as a benchmark value for verifying the performance of the extreme value methods. We observe that both extreme value approaches show a good correspondence with the empirical 99\% quantile for $k \in [150,300]$. Given the similar values of $\hat{\alpha}$ and $\hat{\alpha}^{(c)}$, the two quantile estimators (and their confidence intervals) are also similar.

Besides the full sample result, we also estimate the 99\% quantile for rolling windows spanning eight years of data each, rolling the estimation windows every year. This results in 54 rolling windows, each consisting of approximately $2000$ datapoints. We use both the standard Hill estimator and the bias-corrected Hill estimator, where in the latter estimator, we again set the second order parameter at $-1$.

\begin{figure}[hp]
    \centering
    \includegraphics[width=0.6\textwidth]{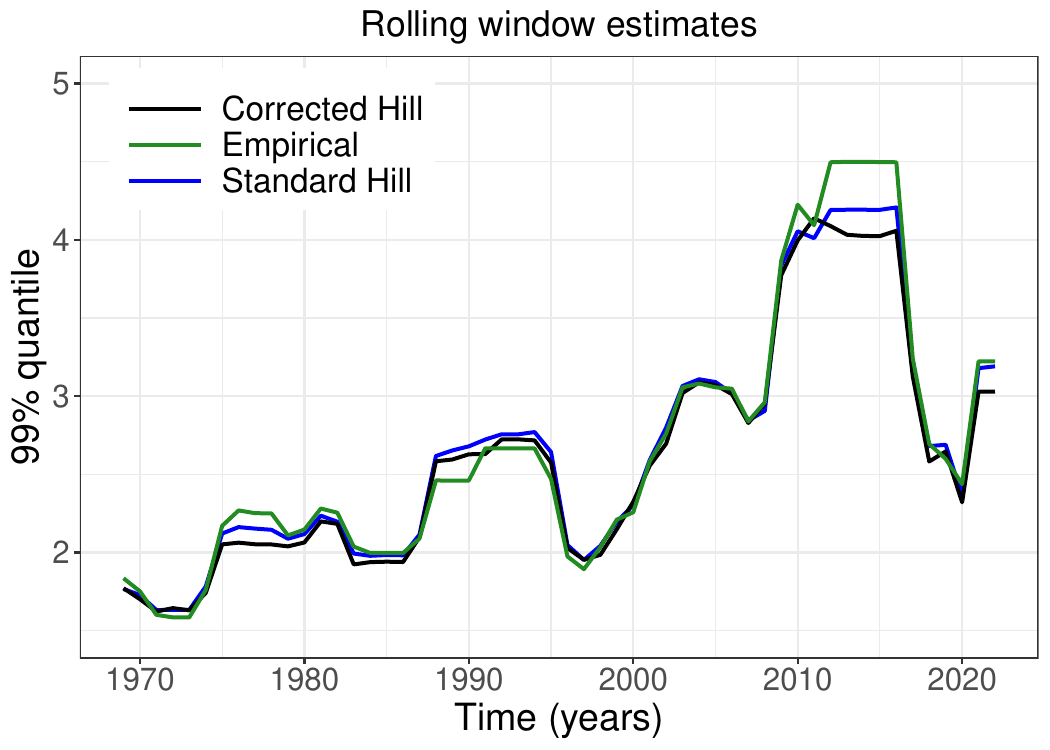} 
    \includegraphics[width=0.6\textwidth]{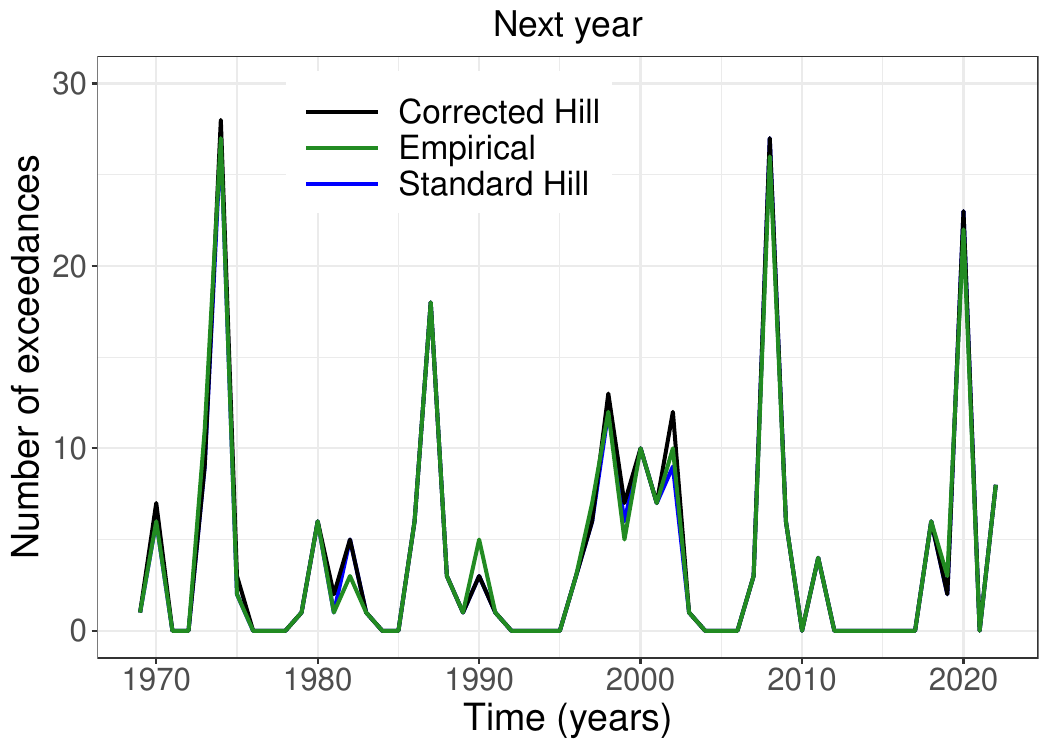} 
    \includegraphics[width=0.6\textwidth]{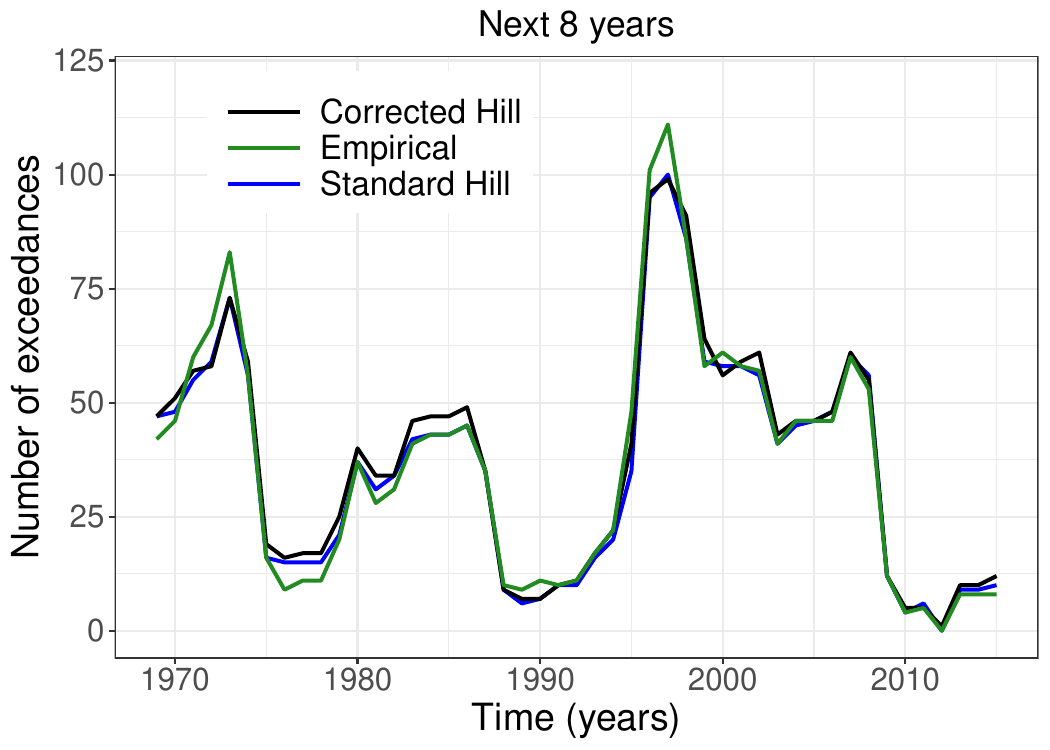} 
    \caption{For the 99\% quantile: rolling window estimates (8 years of data, rolled one year each time), number of exceedances in the next year (expected: $2.5$), number of exceedances in the next 8 years (expected: $20$).}
    \label{fig:sp500roll}
    \end{figure}

From the analysis using the full dataset, we observed that $k/n \approx 0.015$ led to good results for the standard Hill estimator, while $k/n \approx 0.06$ was appropriate for the bias-corrected Hill estimator. With this consideration, we take $k_{\alpha} = 50$ for the standard Hill estimator and $k_{\alpha} = 200$ for the bias-corrected Hill estimator in the rolling window approach. We then estimate $F^{-1} (0.99)$ choosing $k = 50$. Figure \ref{fig:sp500roll} (top) shows that all quantile estimates are in good correspondence with the empirical quantile, with some deviations between 2010 and 2015. Given that the rolling window sample size is approximately 2000, the empirical quantile might not be a very reliable estimate for the high quantile. 

Taking the point estimate of the high quantile as a risk forecast, we count the number of exceedances beyond the estimated quantiles in the next year. The average yearly number of exceedances are $4.15$, $4.31$, and $4.19$, for the standard Hill, the bias-corrected Hill, and the empirical estimator respectively; see also Figure \ref{fig:sp500roll} (middle). These are all higher than the expected value of $2.5$ (corresponding to 250 trading days per year). The reason could be due to several outliers. For example, the 99\% quantiles estimated from the data before the global financial crisis (2000--2007) are exceeded 27 times in 2008, for both the standard and bias-corrected Hill estimators (and 26 times for the empirical estimator). In addition, it could also be due to the fact that the testing observations from only one year may not be sufficient in representing the long-run stationary distribution.

To investigate the validity of the predicted quantile in a longer horizon, Figure \ref{fig:sp500roll} (bottom) shows the number of exceedances beyond the estimated quantiles for the three estimators in the next 8-year window. The average numbers of exceedances are $36.2$, $37.8$, and $36.8$, for the standard Hill, the bias-corrected Hill, and the empirical estimator respectively, much high than the expected value 20; see also Figure \ref{fig:sp500roll} (bottom). 
These numbers are even higher than 8 times the average yearly exceedance numbers. However, there are less outliers: while there was an occasion with 28 yearly excesses for the 1-year testing window, the maximum number of exceedances for the 8-year testing window is $111$, about $111/8 \approx 14$ per year. This suggests that the stationary distribution assumption is potentially violated in such a long estimation and testing window (16 years in total).

To summarize, when analyzing the original dataset without time series filtering, extreme value methods yield similar estimates as the empirical method. The estimated quantile is not valid in a short testing window in the future. Although the estimated quantile seems more robust for a longer testing window, it still largely deviates from the expected performance. Our result shows that the presence of serial dependence can affect the estimation for the high quantile of the stationary distribution.

\section{Conditional Risk Analysis}\label{sec:conditional}
%\ak{Please clarify how the results of this section relate specifically to the S\&P 500 data.}
%\cz{I find it a bit difficult: we cannot say that these results hold for all stocks etc. The purpose of the chapter is to provide a showcase, not a unified conclusion. For a different dataset, one can follows the same "procedure", but not for sure getting the same "conclusion".}
In this section, we consider conditional risk forecasts as follows. We first fit a time series model, namely the  AR(1)--GARCH(1,1) model to the dataset, as described in Section 1.2.2. Then we filter out the estimated residuals $\hat\varepsilon_1,\ldots,\hat\varepsilon_n$. Next, we investigate the estimated residuals as the ``dataset'' and perform extreme value analysis based on those observations.

\begin{figure}[ht]
    \centering
    \includegraphics[width=0.7\textwidth]{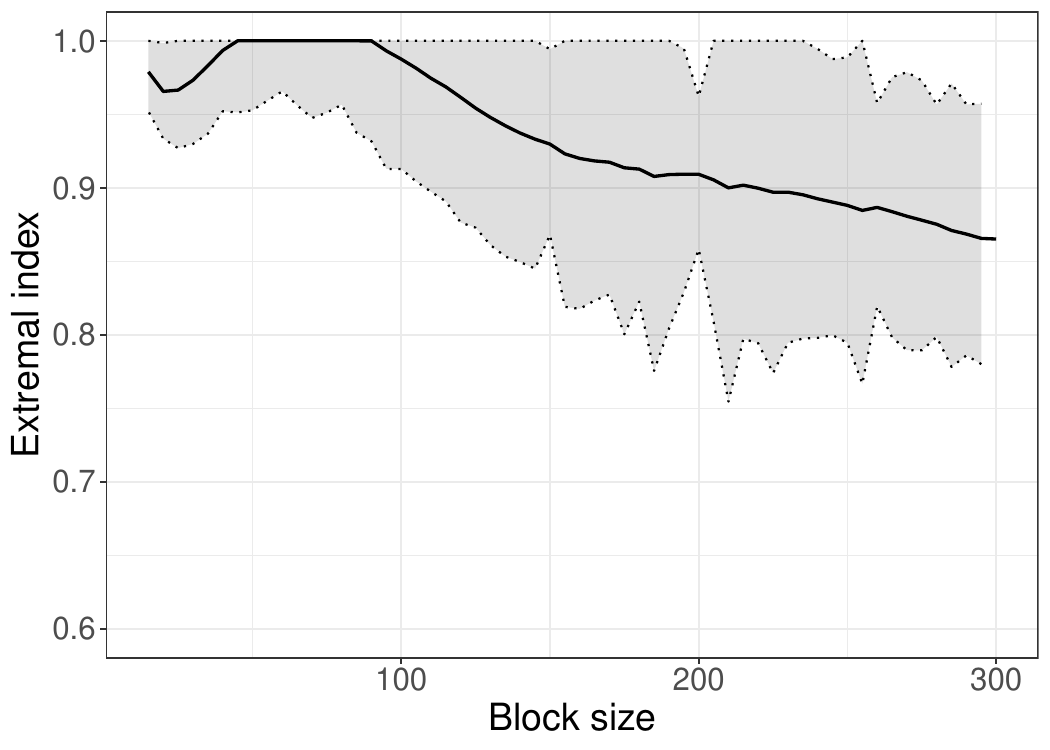} 
    \caption{Bias-corrected sliding blocks estimator, with 95\% confidence intervals, for the residuals of the fitted AR(1)--GARCH(1,1) model.}
    \label{fig:sp500thetaresid}
    \end{figure}

We start by estimating the extremal index $\theta$ of the filtered residuals, using again the bias-corrected sliding blocks estimator \cite{berghaus2018weak}. Figure \ref{fig:sp500thetaresid} shows that we do not reject an extreme index of $1$ for a large range of block size, leading to the conclusion that serial dependence in the extremes is no longer present in the filtered residuals. We remark that although the time series filtering was aiming at filtering out serial dependence in general, it also achieved the same goal for extremes.

\begin{figure}[ht]
    \centering
    \includegraphics[width=0.45\textwidth]{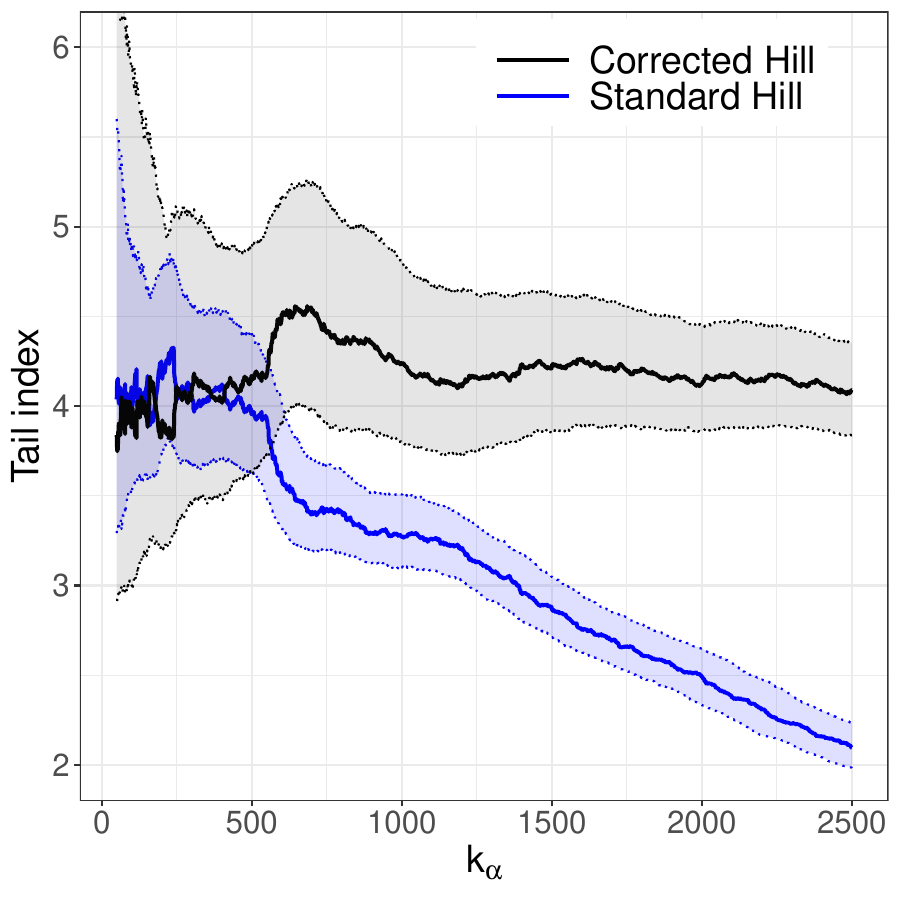} 
    \includegraphics[width=0.45\textwidth]{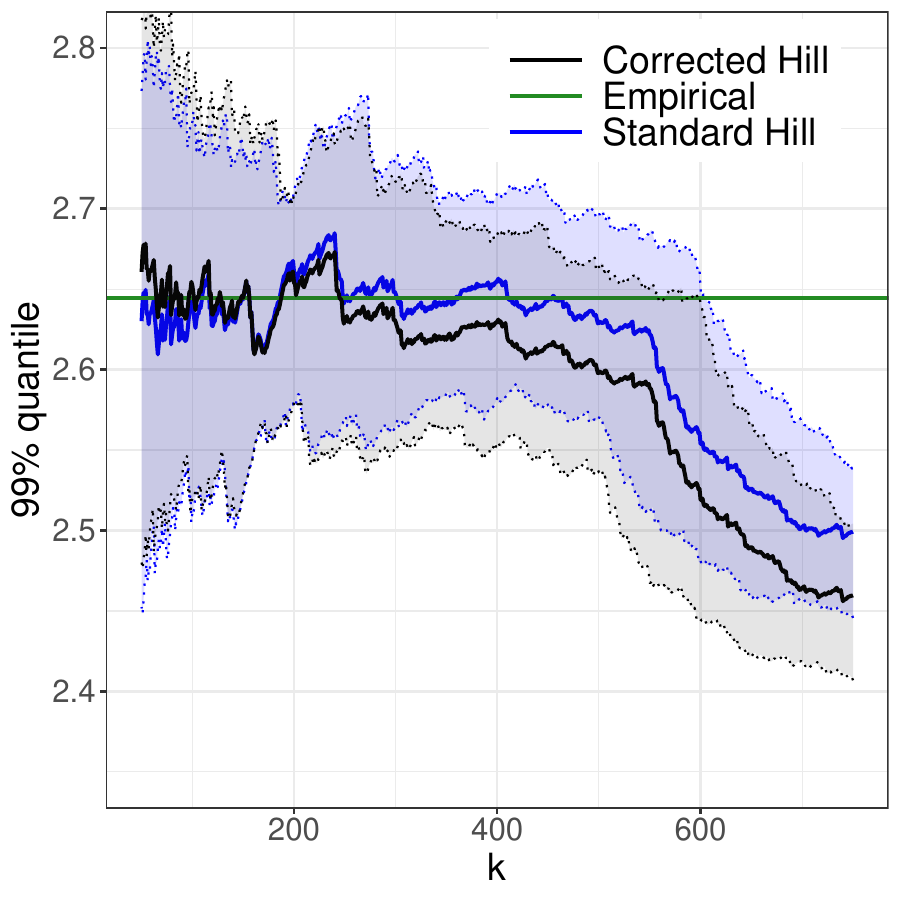} 
    \caption{Estimates of the tail index $\alpha$ and the 99\% quantile based on the standard Hill estimator (blue line) and bias-corrected Hill estimator (black line) for the of the filtered residuals of the daily negative log-returns of the S\&P 500. The horizontal green line represents the empirical 99\% quantile.}
    \label{fig:sp500garchhill}
\end{figure}

Next, we estimate the tail index $\alpha$ and the $99\%$ quantile for the full dataset of residuals. The goal is to compare the standard Hill estimator with the bias-corrected Hill estimator \citep{de2016adapting} (using again $\rho = -1$) 
and verify whether similar values of $k$ and $k_{\alpha}$ as in Section~\ref{sec:unconditional} are appropriate. 
Figure \ref{fig:sp500garchhill} (left) shows that one should choose similar $k$ values for the standard Hill estimator ($k_{\alpha} = 250$) and slightly higher values for the bias-corrected Hill estimator ($k_{\alpha} = 1500$). With these choices, the estimates of the tail index for the residuals are higher than for the original data: $\hat{\alpha} = 4.05$ and $\hat{\alpha}^{(c)} = 4.22$. This confirms the theoretical result that the stationary distribution of a GARCH(1,1) model has a heavier tail than the distribution of the residuals; see \cite{mikosch2000limit}. Concerning the 99\% quantile of the residuals, the empirical estimate is $2.64$. Figure \ref{fig:sp500garchhill} (right) shows that both estimators show good correspondence for $k \in [50,300]$.

Then, we focus on predicting the next-day 99\% quantile using a window spanning eight years of data and rolling this estimation window each day. More specifically, we start by fitting an AR(1)--GARCH(1,1) model to an estimation window consisting of $n_w = 2000$ observations, extracting the residuals, and then forecasting the day-ahead 99\% quantile. For the estimation of the quantile of the residual, we use the quantile estimator \eqref{eq:quant1} with $k = 50$ and plugging in either the standard Hill estimator with $k_{\alpha} = 50$ or the bias-corrected Hill estimator with $k_{\alpha} = 200$. In addition, we also obtain the empirical 99 \% quantile from the filtered residuals. To obtain a final forecast for the next-day 99\% quantile, we combine the estimated quantiles of the residual with the forecasted $\widehat{\sigma}_t$ and $\widehat{\mu}_t$ estimated for the next day based on the estimated time series model. 

For each window of size $n_w = 2000$, we apply the forecasting procedure described above to predict the 99\% for the next day. By comparing the prediction with the realized loss, we obtain an indicator indicating whether the realized loss exceeds the predicted quantile. After shifting the window one day, we repeat the same procedure to obtain a new prediction and consequently a new indicator. By collecting such indicators, we can evaluate the performance of the underlying quantile forecasting method. In particular, we aggregate the indicators in a testing window consisting of $n_{\textnormal{test}} = 250$ days (roughly a year) and shift such a testing window per day.

\begin{figure}[ht]
    \centering
    \includegraphics[width=0.45\textwidth]{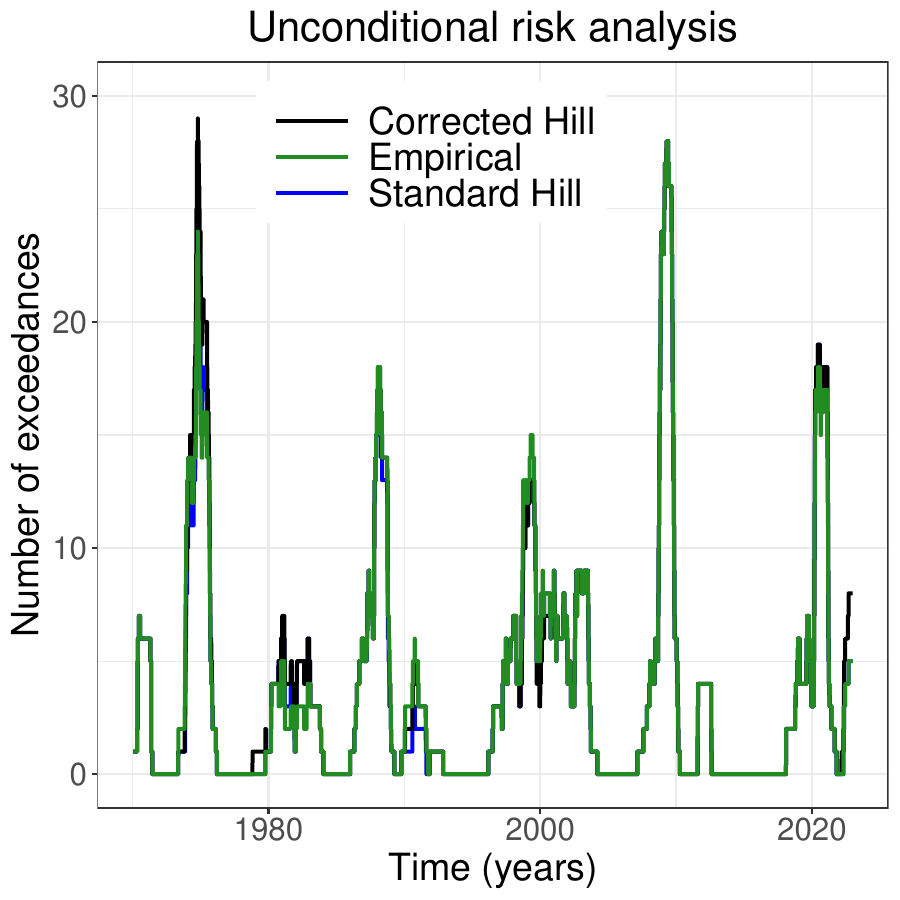} 
    \includegraphics[width=0.45\textwidth]{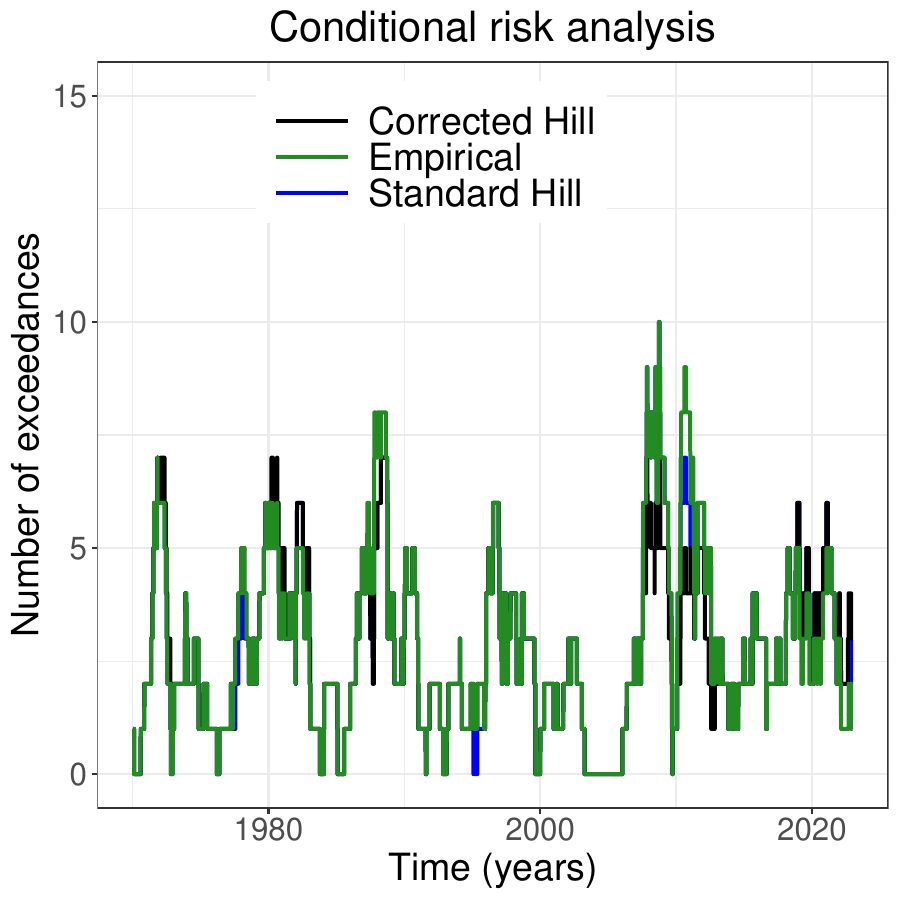}  \\
    \includegraphics[width=0.45\textwidth]{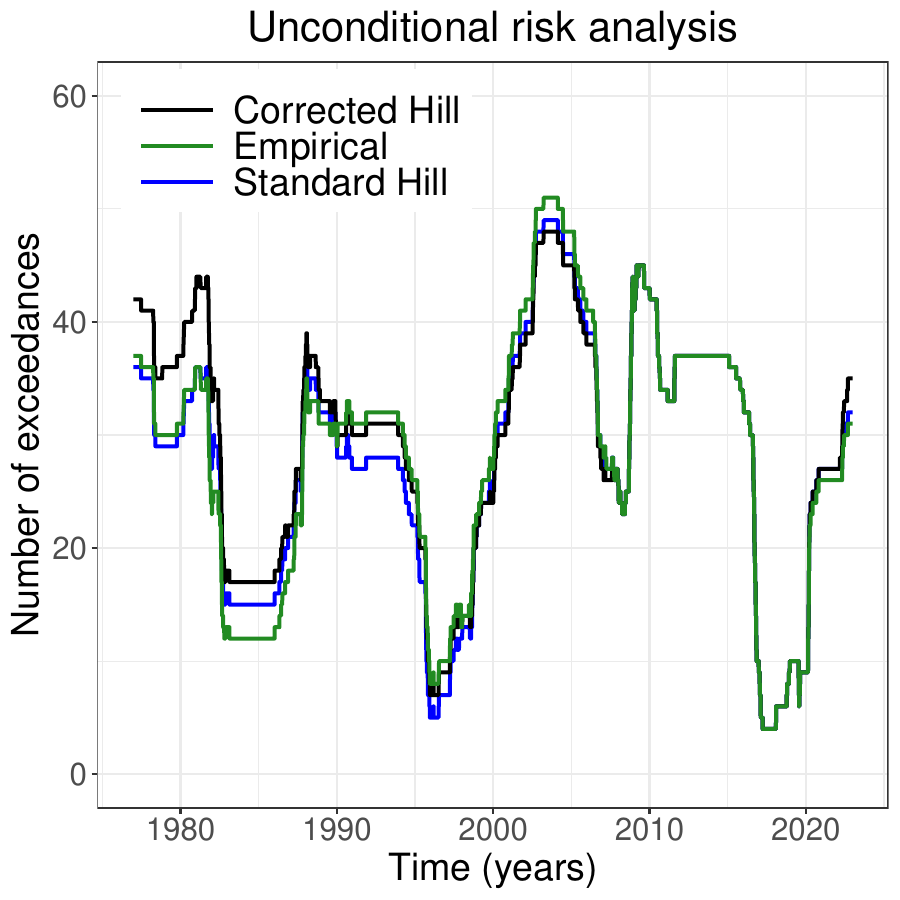} 
\includegraphics[width=0.45\textwidth]{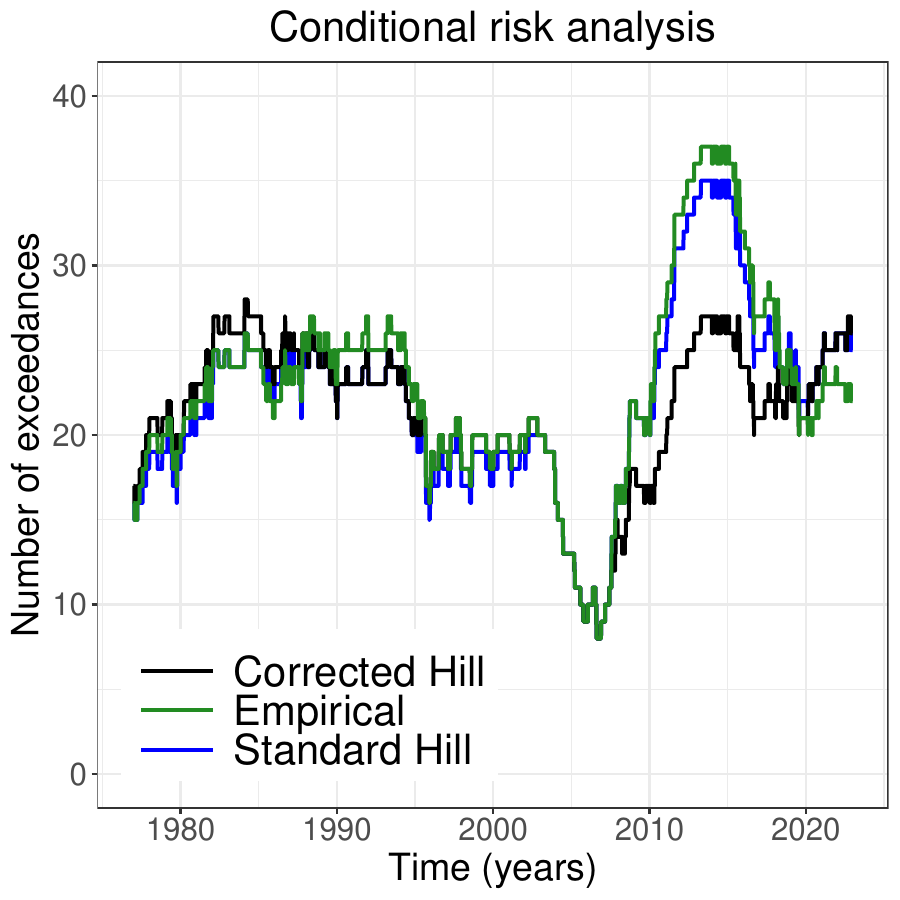} 
    \caption{Rolling window results: the number of exceedances of the 99\% quantile for the unconditional and conditional risk analyses for a testing window of length 250 (top) and of length 2000 (bottom).}
    \label{fig:sp500garchroll}
    \end{figure}

The two top panels of Figure~\ref{fig:sp500garchroll} show the aggregated number of exceedances. For the sake of comparison, the left panel corresponds to the result based on an unconditional analysis using the original dataset without time series model filtering, in the top panel. This figure is similar to the middle panel of Figure \ref{fig:sp500roll}. However, the result is ``finer'' due to daily shifting of the testing window instead of yearly shifting. The results based on the conditional risk analysis are shown in the right panel. We find an average number of exceedances of $3.64$ (standard Hill estimator), $3.86$ (bias-corrected Hill estimator), and $3.70$ (empirical quantile) for the unconditional approach, higher than the expected value $2.5$. The conditional risk analysis performs much better, leading to an average number of exceedances of $2.78$ (standard Hill estimator), $2.71$ (bias-corrected Hill estimator) and $2.81$ (empirical quantile), close to the expected value.

In addition, we use a longer testing window consisting of the next $n_{\textnormal{test}} = 2000$ days (roughly 8 years). The two bottom panels of Figure~\ref{fig:sp500garchroll} show the number of exceedances for the unconditional risk analysis using the original dataset (left) and the conditional risk analysis based on time series model filtering (right). For the unconditional risk analysis, we find an average number of exceedances of $27.71$ (standard Hill estimator), $29.13$ (bias-corrected Hill estimator) and $28.23$ (empirical quantile), higher than the expected value $20$. Again, the conditional risk analysis improves the performance, leading to an average number of exceedances of $22.13$ (standard Hill estimator), $21.35$ (bias-corrected Hill estimator) and $22.83$ (empirical quantile). In addition, we observe less outliers for the conditional risk analysis: the maximum number of yearly exceedances is 29 for the unconditional risk analysis, but is lowered to 10 for the conditional risk analysis. When using a 8-year testing window, the result is similar, the maximum number of 8-year exceedances is 51 for the unconditional risk analysis, which is lowered to 37 when using the conditional risk analysis. Most notably, the bias-corrected Hill estimator performs best among all methods employed.

\section{Backtesting}\label{sec:backtesting}
While counting the number of exceedances in a testing period provides an indication of the performance of a risk analysis method, we would also like to formally test the validity of such a method. For that purpose, we resort to the so-called \textit{backtesting} of risk measures, particularly concerning a predicted high quantile. The idea is to account for the potential randomness in the occurrence of exceedances, while formally testing the null hypothesis that the underlying risk analysis method yields correct forecasts for the quantile.

The essential idea is as follows: under the null hypothesis that a risk analysis method accurately predicts the high quantile, the indicator corresponding to whether a new observation exceeds the forecasted quantile follows a Bernoulli distribution, with probability level equal to the quantile. %For instance, to forecast 99\% quantile, the probability of exceeding is 1\%. 
In addition, under the assumption that such exceedances are independent over time, one can aggregate the number of exceedances within a testing period, which then follows a binomial distribution. The null hypothesis will be rejected if the realized number of exceedances deviates too far from what is expected from the binomial distribution. %Hence, this testing procedure mainly focuses on comparing the exceedance probability with the intended probability level.

In this section, we consider two classic tests: the unconditional coverage (UC) test and conditional coverage (CC) test. The UC test focuses on testing whether the exceedance probability matches the intended probability level, as described above. %, similar to the \cite{kupiec1995techniques} test. 
By contrast, the CC test aims to verify whether the exceedances are independent over time. The details of the two tests are as follows. For a testing period with sample size $n$, let $n_1$ denote the number of exceedances, and define $\pi=n_1/n$ and $n_0=n-n_1$. The UC test statistic is actually a likelihood ratio test based on the Bernoulli likelihood: as $n\to\infty$, under the null hypothesis of correct unconditional coverage,
\[
 \textnormal{LR}_{\textnormal{uc}}  = -2\log\frac{(1-p)^{n_0}p^{n_1}}{(1-\pi)^{n_0}\pi^{n_1}} \stackrel{\text{d}}{\to} \chi^2_1,
\]
where $p$ is the intended exceedance probability.

For the CC test, to test the independence among exceedances, the idea is to consider the observed exceedance indicators (Bernoulli random variables) as a Markov process. Then we have the following likelihood ratio test based on the independence transition matrix: as $n\to\infty$, under the null hypothesis of independent coverage,
		\[
  \textnormal{LR}_{\textnormal{ind}} = -2\log\frac{(1-\pi)^{n_{0,0}+n_{1,0}}\pi^{n_{0,1}+n_{1,1}}}{(1-\pi_{0,1})^{n_{0,0}}\pi_{0,1}^{n_{0,1}}(1-\pi_{1,1})^{n_{1,0}}\pi_{1,1}^{n_{1,1}}} \stackrel{d}{\to }\chi^2_1,
  \]
where $n_{i,j}$ counts the number of transitions from $i$ to $j$ in consecutive days for $i,j\in\{0,1\}$, $\pi=(n_{0,1}+n_{1,1})$ is an estimate of the exceedance probability, and $\pi_{i,1}=n_{i,1}/(n_{i,0}+n_{i,1})$ is an estimate of today's conditional exceedance probability given that the previous day was in state $i$ for $i\in\{0,1\}$. One can then combine the two likelihood ratio tests to one augmented test of the overall performance of the method because the two test statistics are asymptotically independent: as $n\to\infty$, under the joint null hypothesis of having both correct unconditional coverage and independent coverage,
	
 \[
\textnormal{LR}_{\textnormal{cc}}=\textnormal{LR}_{\textnormal{uc}}+\textnormal{LR}_{\textnormal{ind}} \stackrel{\text{d}}{\to} \chi^2_2.
 \]

The choice of the backtesting method should be in line with the goal of the underlying risk analysis. For instance, for an unconditional risk analysis, the goal is to predict an unconditional quantile for the stationary distribution. As a consequence, the predicted quantile is expected to be a valid quantile for a longer horizon, but there is no guarantee that the exceedances of such an unconditional quantile are not clustered in crisis periods. Consequently, only the UC test should be applied to testing the performance of an unconditional quantile forecast. By contrast, the conditional risk analysis aims at producing dynamic risk forecasts, i.e., a conditional quantile for a short horizon, such as the next day. Due to the conditional feature, the predicted conditional quantile is expected to vary across time, incorporating macroeconomic and/or financial market information. A good conditional risk forecast method should produce  conditional quantile exceedances that not only reflect the intended probability level but also remain independent over time. Consequently, one should apply both the UC and CC tests to validate the performance of a conditional risk forecast method.

We consider backtesting for the three methods of quantile estimation discussed in the previous sections, for both the unconditional (Section~\ref{sec:unconditional}) and the conditional (Section~\ref{sec:conditional}) approach. For a testing window with a size 250 (roughly 1 year), Table~\ref{tab:tests} (top two rows) shows the percentages of rejections of the null hypothesis for the UC and CC tests for a significance level of $\alpha = 0.05$ under the rolling window setup. Note that here we do not suffer from a multiple test issue: by counting the number of rejections in different testing windows, we expect rejections among 5\% of the tests. A lower number is further preferred.

We observe that for the unconditional risk analysis, none of the proposed method reaches the intended 5\% rejection level, both for the UC and the CC tests. Note that applying the CC test here is not really meaningful, but surprisingly also the UC test shows that all methods fail. We argue that this could be due to the reason that the testing window of 250 days is not sufficiently long to represent the stationary distribution. This short horizon effect has been demonstrated by the high number of exceedances in Figure \ref{fig:sp500roll}.

By contrast, for the conditional risk analysis, all three methods yield good performance. While the empirical method yields around 5\% rejection results, the extreme value methods result in lower number of rejection. This shows that among more testing windows, the extreme value methods produce risk forecasts that are valid. This result is confirmed by both the UC and CC tests.

We perform a similar analysis with a longer horizon for backtesting, namely 2000 days (roughly 8 years). Table~\ref{tab:tests} (bottom two rows) shows that the non-stationarity that is inevitable for such a long testing window causes most methods to fail; only the extreme-value approach based on the bias-corrected Hill estimator for the conditional quantile yields good performance.  

\begin{table}[ht]
\centering
\begin{tabular}{lcccccc}
%\midrule
& \multicolumn{3}{c}{Unconditional quantile} & \multicolumn{3}{c}{Conditional quantile}
\\
\cmidrule(lr){2-4}\cmidrule(lr){5-7}
           & Standard & Corrected & Empirical    &Standard & Corrected & Empirical\\
           \midrule
UC   & 0.174 & 0.182 & 0.178 & 0.035 & 0.021 & 0.058 \\
CC & 0.175 & 0.172 & 0.179 & 0.026 & 0.011 & 0.048 \\
\midrule
UC   & 0.573 & 0.688 & 0.679 & 0.148 & 0.051 & 0.162 \\
CC & 0.808 & 0.788 & 0.736 & 0.127 & 0.037 & 0.160 \\
%\bottomrule
\end{tabular}
\caption{Percentage of rejections of null hypothesis for the unconditional coverage (UC) and conditional coverage (CC) tests for a significance level of $\alpha = 0.05$ for a testing window of length 250 (top two rows) and of length 2000 (bottom two rows).}
\label{tab:tests}
\end{table}

\section{The Tail Dependence Coefficient} \label{sec:multivariate}
The tail risk analysis presented so far focuses on a single risk factor, i.e., on the losses of a single stock market index. When considering multiple risk factors, for instance, when analyzing compound extreme events caused by severe losses of multiple stock market indices, we need a multivariate approach that can cope with \textit{tail dependence}, i.e., with the possibility of having joint tail events in different risk factors. 

A typical summary measure of tail dependence, often used as a model diagnosis tool, is the \textit{tail dependence coefficient}. In a bivariate setting, let $X$ and $Y$ denote two risk factors with a joint distribution function $F$ and marginal distributions $F_X$ and $F_Y$. The (upper) tail dependence coefficient is defined as 
\begin{equation}\label{eq:chi}
\chi:=\lim_{p\to 0} p^{-1} \Prob \left\{F_X(X)>1-p,F_Y(Y)>1-p\right\}.
\end{equation}
The tail dependence coefficient only concerns the dependence structure between $X$ and $Y$, in other words, the distribution of $\left(F_X(X),F_Y(Y)\right)$, also known as the \textit{copula} of $(X,Y)$. For more details about the tail dependence coefficient and its relation to multivariate extreme value theory, see Chapter~7. 
%\cz{I also wonder whether we need the paragraph in Intro to discuss the relation of our chapter with others. We could also move the texts here (and other places for the corresponding chapter).}

The tail dependence coefficient takes values in $[0,1]$. When $\chi>0$, $X$ and $Y$ are called \textit{asymptotically dependent} or \textit{tail dependent}, while $\chi = 0$ corresponds to \textit{asymptotic independence} or \textit{tail independence}. The tail dependence coefficient can be interpreted as the (conditional) probability that $X$ is ``at risk'' given that $Y$ is ``at risk'', or vice versa. Here, ``at risk'' refers to having an extreme event with the same tail probability $p$, where $p$ is very small. When $X$ and $Y$ are losses in financial assets, this interpretation can be regarded as asset market linkages in extreme events; see, e.g. \cite{hartmann2004asset}. In addition, this interpretation can be easily connected with characterizing the presence of systemic risk when considering $X$ and $Y$ as losses in financial institutions; see e.g. \cite{de2010back}.

Suppose now that we have i.i.d. observations $(X_1,Y_1), \ldots, (X_n,Y_n)$ from the (unknown) distribution $F$. We can estimate the tail dependence coefficient as follows.  Let $R_{i}^X$ denote the rank of $X_i$ among $X_1,\ldots,X_n$, and $R_{i}^Y$ denote the rank of $Y_i$ among $Y_1,\ldots,Y_n$. Let $k = k_n$ be an intermediate sequence (as in univariate risk analysis), i.e., $k/n\to 0$ and $k \to \infty$ as $n\to\infty$. Setting $p = k/n$ in \eqref{eq:chi}, an empirical estimate of $\chi$ is obtained by 
	\[
	\hat{\chi} = \frac{1}{k} \sum_{i=1}^n  \mathbb{I} \left( R^X_i > n  - k, R^X_i > n - k \right).
	\]
When assessing the tail dependence of a bivariate financial dataset, 
the same concern as in the univariate context arises: would the serial dependence in the financial data affect the estimation of the tail dependence coefficient? Is it necessary to remove serial dependence, e.g., by applying a time series model %e.g., via GARCH filtering, 
and estimating the tail dependence coefficient based on the residuals extracted from such a model? 

To illustrate the impact of serial dependence on the estimation of the tail dependence coefficient, we take the daily adjusted closing prices of the S\&P 500 index, the FTSE100 index and the Dow Jones Industrial Average (DJIA) from Yahoo Finance, for the period January 1st, 1992 up to December 31st, 2022, and calculate their negative daily log-returns. We then construct two bivariate samples, selecting only days for which both indices are available, (i) the S\&P 500 and DJIA returns, $n = 7808$ and (ii) the S\&P 500 and FTSE100 returns, $n = 7577$.  

\begin{figure}[ht]
    \centering
    \includegraphics[width=0.4\textwidth]{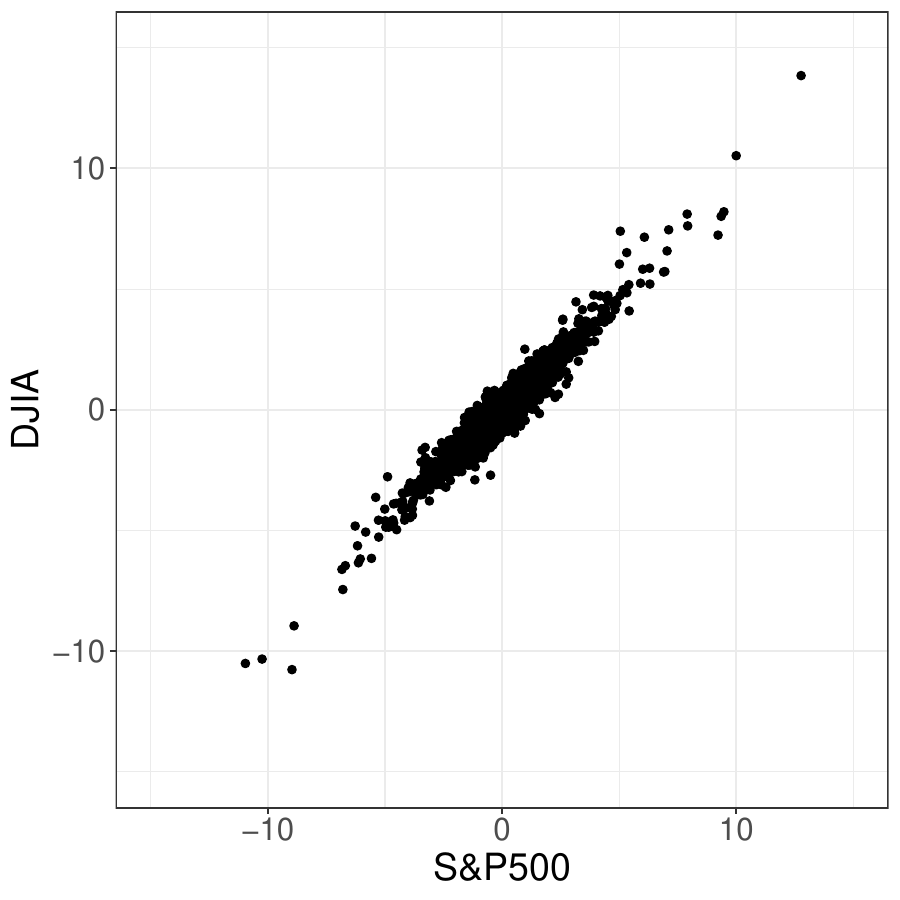} 
    \includegraphics[width=0.4\textwidth]{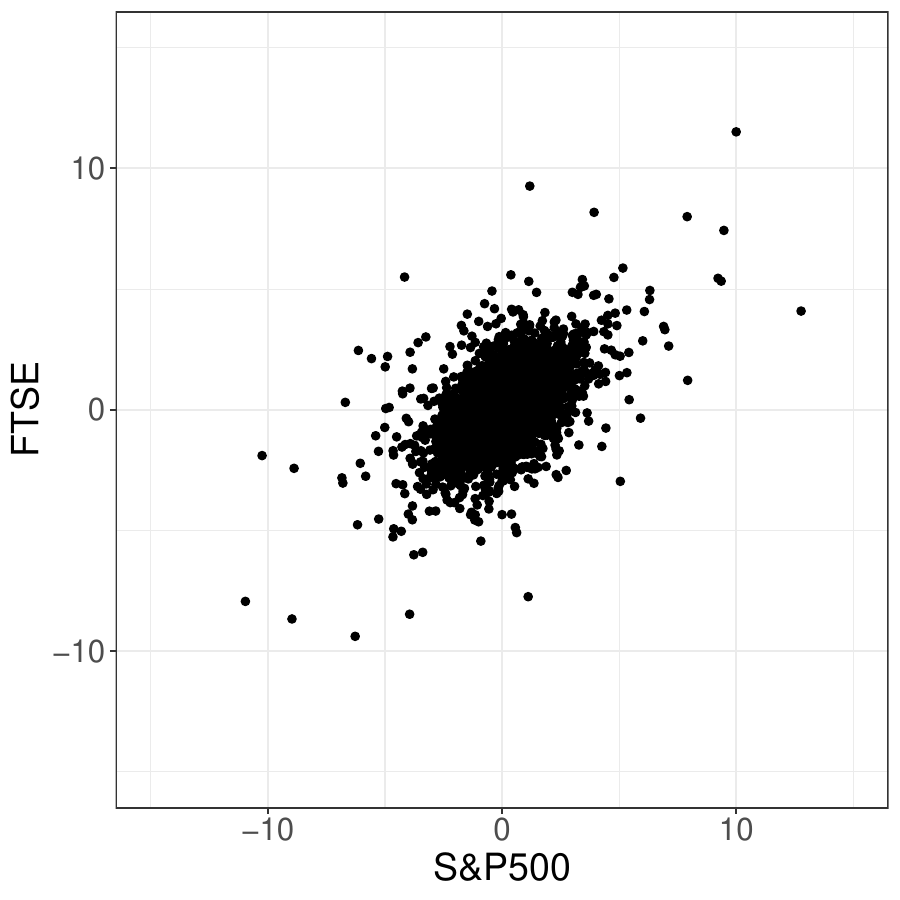}   \\
      \includegraphics[width=0.4\textwidth]{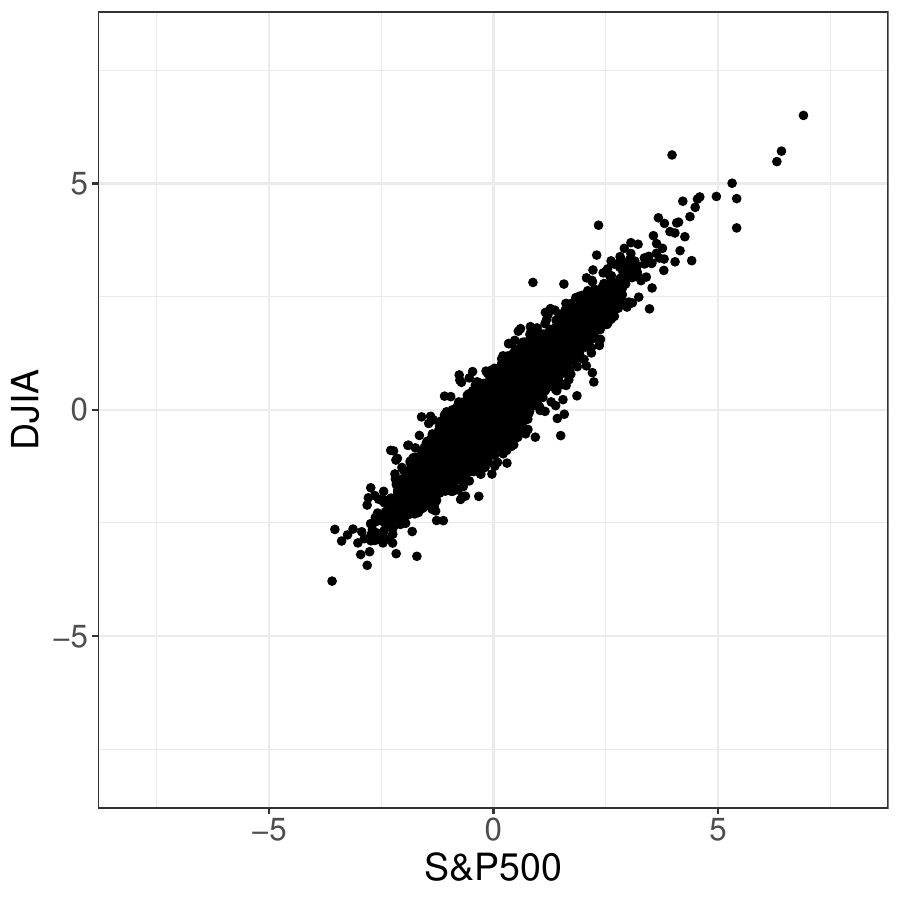}  
    \includegraphics[width=0.4\textwidth]{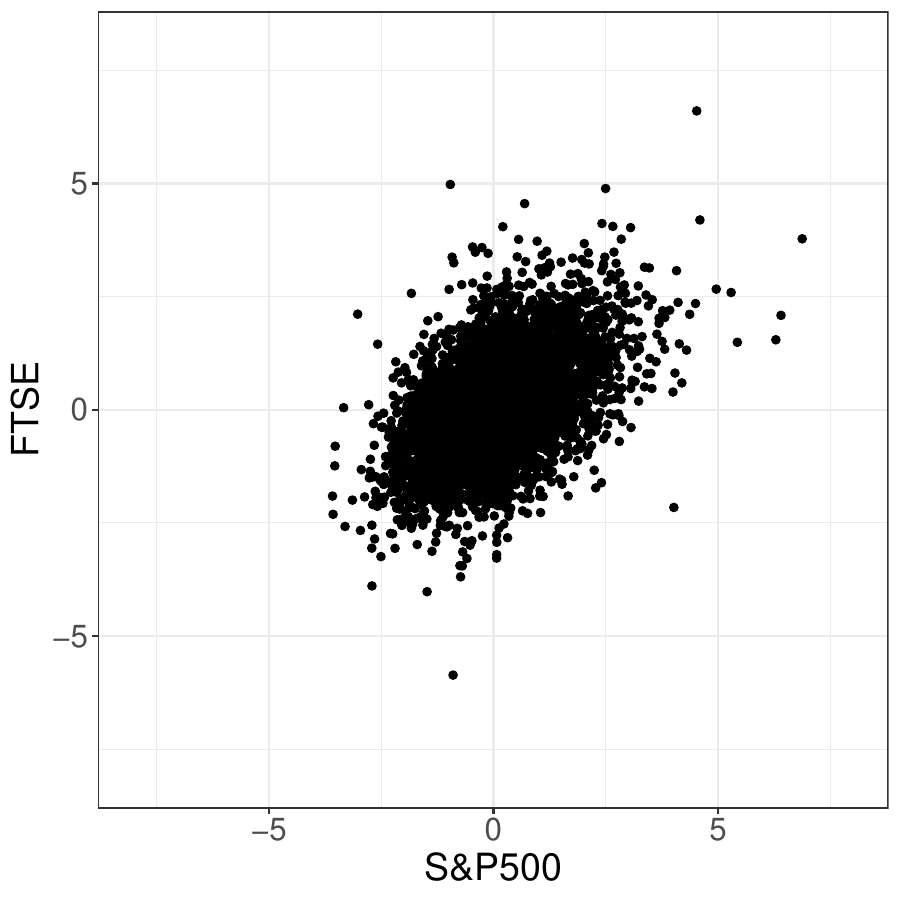}  
    \caption{Scatterplots of the daily negative log-returns of the S\&P 500 versus the DJIA (left) and the S\&P 500 versus the FTSE100 (right), for the raw data (left) and the AR(1)--GARCH(1,1) residuals (right).}
    \label{fig:scatterplots}
    \end{figure}

Figure~\ref{fig:scatterplots} (top) shows scatterplots for these two bivariate datasets. Unsurprisingly, we see much stronger tail dependence between the S\&P 500 and the DJIA than between the S\&P 500 and the FTSE100. 
Next, we apply the AR(1)--GARCH(1,1) model to the three univariate series of log-returns and extract the residuals. Figure~\ref{fig:scatterplots} (bottom) shows scatterplots of residuals of S\&P 500 vs DJIA and of S\&P 500 vs FTSE100. We observe a similar level of bivariate tail dependence as that of the original data.

\begin{figure}[ht]
    \centering
    \includegraphics[width=0.45\textwidth]{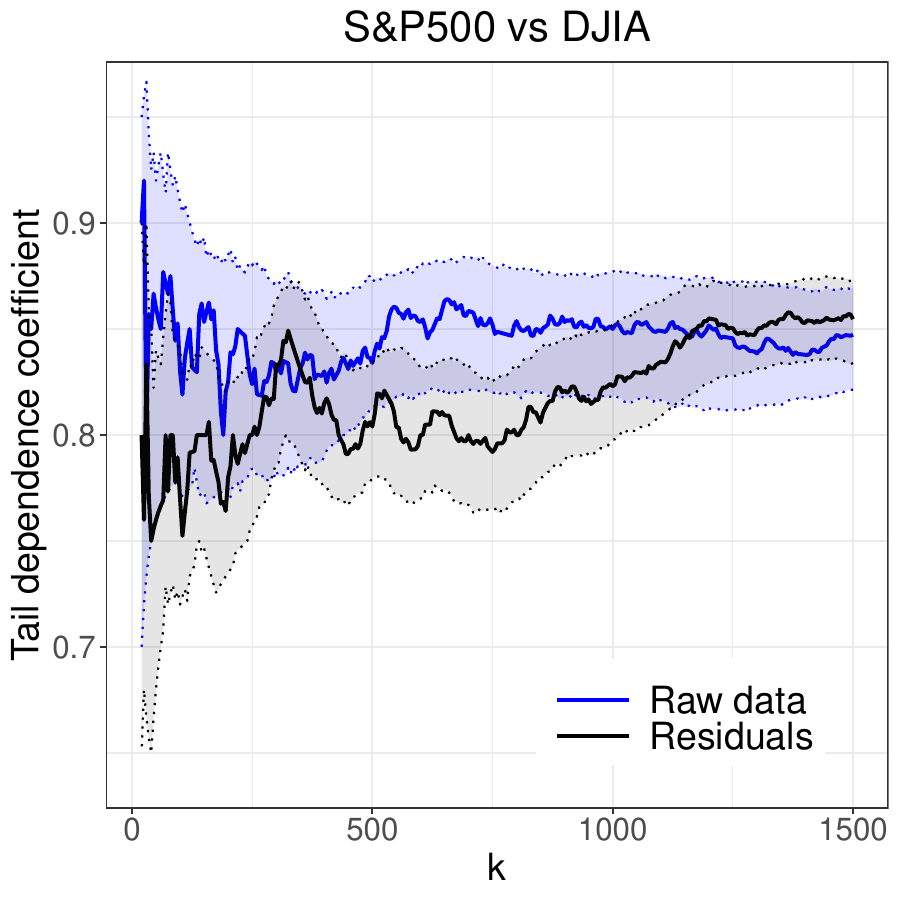}  
\includegraphics[width=0.45\textwidth]{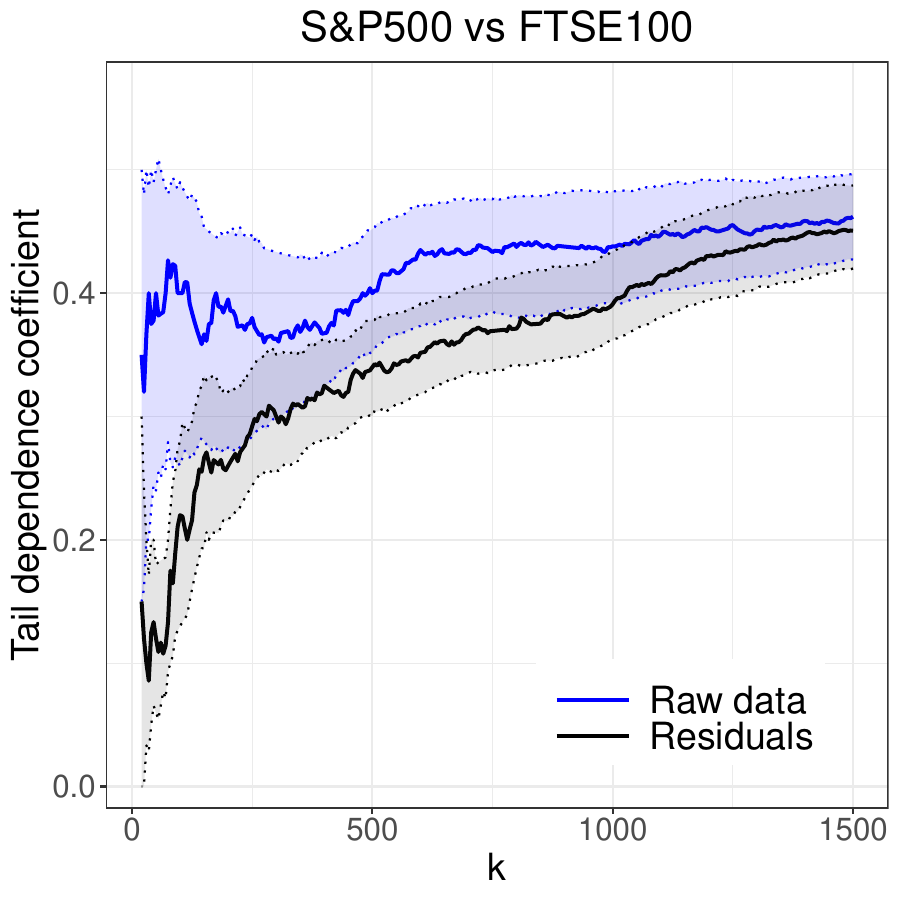}  
    \caption{Estimates of the tail dependence coefficient $\chi$ for the raw data and the residuals of the daily negative log-returns.}
    \label{fig:chi}
    \end{figure}

Figure~\ref{fig:chi} shows estimates $\hat{\chi}$ for the original data and for the residuals. As expected, tail dependence is much stronger for S\&P 500 vs DJIA (left) than for S\&P 500 vs FTSE100 (right). Dotted lines show 90\% confidence intervals obtained via a block bootstrap procedure similar to that for the tail index estimation in the univariate context. For the first dataset, S\&P 500 vs DJIA, we observe that choosing $k\in[350, 800]$, which amounts to $k/n\in[0.05,0.1]$, is appropriate for obtaining an estimate for the tail dependence coefficient, both for the raw data and the residuals. For $k=500$, we obtain $\hat{\chi} = 0.83$ $(0.81,0.88)$ for the raw data and $\hat{\chi} = 0.80$ $(0.78, 0.84)$ for the residuals. While we do observe a slightly lower estimate based on the residuals, statistically, the overlapping confidence intervals do not rule out the statement that the two tail dependence coefficients are the same. Economically, from the point estimates, the magnitude of difference, interpreted as conditional probabilities, is marginal. 
For the second dataset, S\&P 500 vs FTSE100, the choice of $k$ is less obvious. Taking again $k=500$, we obtain $\hat{\chi} = 0.40$ $(0.36,0.45)$ for the raw data and $\hat{\chi} = 0.34$ $(0.30, 0.38)$ for the residuals. Even though estimates for the tail dependence coefficient based on the residuals are consistently lower than those based on the raw data, the confidence intervals still overlap (and both estimates rule out asymptotic independence). We therefore conclude that time series filtering does not affect the estimation of the tail dependence coefficient.

\section{Key Takeaways} \label{sec:conclusion}
% \mainmatter
This chapter showcases the application of extreme value methods for estimating and forecasting tail risk measures in financial data. By exploring an example dataset, our analysis reveals that financial data exhibit heavy tails and serial dependence. %Addressing serial dependence presents a significant challenge in applying extreme value statistics to financial data. 
Our empirical analysis yields key insights into managing serial dependence in financial risk analysis, a significant challenge when applying extreme value statistics to financial data.

First and foremost, it is crucial to define the economic objective behind the risk analysis. Contrasting regulatory and institutional perspectives reveal differing goals: establishing robust capital requirements versus incorporating recent information for risk analysis. For the purpose of establishing a stable and robust risk measure accounting for all potential macroeconomic scenarios, an unconditional quantile for the stationary distribution is preferred, provided that the underlying distribution of the financial data is stationary. By contrast, for the purpose of incorporating the most recent information available in risk analysis, quantiles of the conditional distribution of the losses are needed. Such a conditional risk analysis is crucial for day-to-day operations and strategic decision-making.

Second, estimating unconditional quantiles precludes the use of standard declustering methods due to their inadequacy in eliminating serial dependence in financial extremes. When applying such declustering procedures, serial dependence in financial extremes can be reduced but not eliminated, at the costs of losing a large fraction of data. This finding underscores the limitations of conventional declustering methods in financial contexts. While the practical advice of abandoning declustering methods for financial data is vital, the theoretical reason behind this is yet to be further investigated. 

Third, for unconditional quantile estimation, utilizing the original dataset while acknowledging the presence of serial dependence is the remaining option. However, such an approach compromises short-term reliability. This is due to the fact that any upcoming year may represent a specific economic scenario which deviates from the long run stationary distribution. The produced risk forecast often fails backtesting in a longer testing horizon (e.g., 8-year) as well. This can be explained by the non-stationarity in such a long estimation and testing window (16 years in total).

Fourth, predicting conditional quantiles through a two-step approach yields promising results. The two-step approach first fits the data to a financial time series model, followed by the application of extreme value statistics to the residuals. The validity of such an approach is confirmed by existing backtesting methods. The best performing method is the high quantile estimator in extreme value statistics coupled with the bias-corrected Hill estimator.

Finally, for financial data, the serial dependence in marginals seems to have little or no impact on statistical analysis regarding the cross-sectional tail dependence. This is an encouraging message for systemic risk analysis: using the original time series or the more serially independent residuals after filtering by a time series model yield very similar results at least for the estimation of the tail dependence coefficient. Therefore, one can safely apply multivariate extreme value statistics to financial data, if the target quantity concerns the cross-sectional tail dependence only.

\section{Notes and Comments}
\label{sec:notescomments}

This chapter focuses on the main challenge of applying extreme value statistics to financial data: handling serial dependence. To obtain an extensive and systematic understanding concerning financial extremes, we refer the readers to early literature on the fundamentals of extreme value theory and its applications in finance and insurance; for example, see the textbooks \cite{embrechts2013modelling} and \cite{longin2016extreme}.

A key tool used for examining the presence of serial dependence is the extremal index. Estimation of the extremal index has been widely studied in the literature; see, e.g. \cite{smith1994estimating} and \cite{ferro2003inference}. More recently, \cite{northrop2015} proposed a sliding blocks estimator, which was analyzed theoretically in \cite{berghaus2018weak}. 
%\cite{berghaus2018weak} shows the asymptotic normality of this estimator under mild conditions, while 
Finally, \cite{bucher2020} proposes further refinements that have smaller variance when $\theta$ is close to 1. 

When estimation of an unconditional high quantile is of interest, we recommend to apply the Weissman estimator while acknowledging the presence of serial dependence. The theoretical guarantee of this approach can be found in \cite{drees2000weighted} and \cite{drees2003extreme}. Under mild conditions for serial dependence, the Weismann estimator possesses asymptotic normality with the same asymptotic bias as in the i.i.d. case, but a different variance related to the serial dependence. 

We also discussed backtesting for risk forecasts. Earlier backtesting methods date back to \cite{kupiec1995techniques}. Following the same ideas, \cite{christoffersen1998evaluating} proposed two tests: the unconditional coverage (UC) test and conditional coverage (CC) test. Other tests on the independence of exceedances can be found in \cite{bucher2020using}. In this chapter we did not consider comparing the performance between different risk analysis methods. To achieve that goal, one can use comparative tests; see e.g., \cite{nolde2017elicitability}.

While the focus of this chapter is mainly on univariate approaches, we briefly touched up on multivariate extreme value statistics in financial applications. Estimation of a generalization of $\chi$ (the so-called \textit{tail copula}) has been analyzed theoretically in \cite{schmidt2006non}, among others. To the best of our knowledge, the theoretical properties of the estimate of $\chi$ under serial dependence have yet to be investigated. For further extensive examples of applying multivariate extreme-value statistics in finance, see Chapter~8 as well as \cite{longin2001extreme,poon2004extreme}and \cite{castro2018time}.

\bibliographystyle{plainnat}
\bibliography{bibtex_example}

\end{document}